# Ultrafast electron diffractive imaging of the dissociation of pre-excited molecules

Yanwei Xiong*, Haoran Zhao, Sri Bhavya Muvva, Cuong Le, Lauren F. Heald, Jackson Lederer, and Martin Centurion†

Department of Physics and Astronomy, University of Nebraska-Lincoln, Lincoln, Nebraska, 68588, USA

**ABSTRACT**. Gas phase ultrafast electron diffraction (GUED) has become a powerful technique to directly observe the structural dynamics of photoexcited molecules. GUED reveals information about the nuclear motions that is complementary to the information on the electronic states provided by spectroscopic measurements. GUED experiments so far have utilized a single laser pulse to excite the molecules and an electron pulse to probe the dynamics. This limits the excited states which can be studied to only those that can be reached by absorption of a photon from the ground state or in some cases simultaneous absorption of multiple photons. A broader class of experiments and dynamics can be accessed using two time-delayed laser pulses to access unexplored regions of the potential energy surfaces. As a proof-of-principle experiment using a double excitation, we studied the photodissociation of trifluoroiodomethane molecules that are pre-excited with an infrared (800 nm) femtosecond laser pulse before photo-dissociation is triggered with an ultraviolet (266 nm) femtosecond laser pulse. We have observed significant differences in the dissociation dynamics, with pre-excitation resulting in a slower dissociation process. This new capability can offer new insights on the evolution of nuclear wavepackets in regions of the excited potential energy surface which are inaccessible in single photon excitation. We present a methodology to carry out the measurement, analyze and interpret the data that could be applied to a broad class of experiments.

## I. INTRODUCTION

Gas phase ultrafast electron diffraction (GUED) has become a powerful tool to determine the nuclear motions and relaxation dynamics of photoexcited molecules [1-6]. GUED is directly sensitive to the nuclear geometry of the molecules and thus complementary to spectroscopic measurements that capture changes in the electronic state. GUED provides high spatial and temporal resolution, and the diffraction observable is straightforward to calculate, making it an ideal tool to capture coherent structural changes and to benchmark theoretical calculations of molecular dynamics. The early GUED work of the Zewail group demonstrated the capability to observe transient states and reaction kinetics with picosecond temporal resolution [7, 8]. The next milestone was the direct observation of nuclear motions during a photochemical reaction, which required femtosecond resolution. The use of radio-frequency (RF) accelerators to deliver relativistic electrons with mega-electron-volt (MeV) kinetic energy has enabled GUED to achieve a temporal resolution of 150 fs [9-11], and more recently less than 100 fs [12, 13]. Through the use of RF compression cavities [14, 15] and optical compensation of the electron-laser velocity mismatch, sub-relativistic GUED with kilo-electron-volt (keV) electrons reached a temporal resolution of 240 fs [16, 17]. The femtosecond temporal resolution has enabled major scientific advances such as imaging the splitting of a nuclear wavepacket at a conical intersection [3], the simultaneous observation of changes in the nuclear and electronic configurations [18], capturing coherent nuclear motions in photoexcited molecules after they return to the ground state [19, 20], and following the structural rearrangements in ring opening reactions in real time [21, 22]. Recently, more advanced analysis methods and improvements in data quality have enabled studies on more complex reactions involving larger molecules and multiple reaction channels [23-25]. GUED has also been used to record 3D molecular images and to study rotational wavepackets [16, 17, 26, 27]. More recently, the capabilities of GUED have been extended to investigate dynamics triggered by ionization through the use of *ab-initio* scattering calculations and more advanced data analysis methods [28, 29]. The method has also been extended to studying liquid phase samples [30, 31] resulting on ground breaking studies on the dynamics of liquid water [32]. Thus far, improvements in GUED capabilities have led to compelling scientific advances by revealing key details of the relaxation dynamics or by opening new avenues of research. There remains an important limitation to the type of experiment that can be carried out with GUED, i.e. the use of a single pump laser pulse to prepare the excited state to be studied. Introducing a pair of time-delayed pump laser pulses would enable additional capabilities to prepare the sample and probe the dynamics. For example, two pulses could be used to vibrationally pre-excited the molecules and alter the reaction dynamics and products [33-35], to pre-align the molecules and capture the dynamics in the molecular frame [36], or to populate a new excited state by letting the wavepacket evolve after the first excitation and then transferring the population to a different state, or by using a strong field pulse to alter the dynamics [33, 37].

Here we present a proof-of-principle GUED experiment using a pair of time-delayed femtosecond laser pulses to trigger photodissociation dynamics in trifluoroiodomethane ($CF_3I$) molecules. An IR femtosecond laser pulse is used to pre-excite $CF_3I$ molecules, followed by a UV femtosecond laser pulse that starts a photodissociation reaction, and a femtosecond electron pulse to probe the structural dynamics. We demonstrate the experimental capability and the methodology to retrieve information from changes in both the structure and angular distribution of the molecules. In addition, we demonstrate a time resolution of 150 fs for GUED using a table-top keV electron source, which significantly improves the previous limit for sub-relativistic GUED and matches the temporal resolution of the state-of-the-art relativistic MeV-UED user facility at SLAC National Accelerator Laboratory [9-11].

*Contact author: yxiong3@unl.edu
†Contact author: martin.centurion@unl.edu

The UV excitation of $CF_3I$ to the A-band, which consists of three repulsive states overlapped in energy denoted as $^3Q_1$, $^3Q_0$, and $^1Q_1$, promotes a non-bonding (n) electron from the iodine atom valence shell to the $\sigma^*$ anti-bonding orbital of the C-I bond, resulting in a prompt cleavage of the C-I bond [38-45]. The rapid breaking of the C-I bond of $CF_3I$ molecules happens within 50 fs [3, 10].

We use an intense IR femtosecond laser pulse that produces a ro-vibrational excitation in the electronic ground state of the molecule. Three experiments were carried out to fully characterize the dynamics: using only the UV excitation, using only the IR pulse, and using a double-pulse excitation with IR pulse followed by the UV pulse. We have observed that, when the molecules are pre-excited by the IR laser pulse, the dissociation process is significantly slowed down with respect to the UV-only excitation. We present here the required methodology for carrying out the measurements and analyzing and interpreting the diffraction data that includes multiple excitations and transient anisotropy. This new capability significantly advances the state of the art and opens new research directions for GUED in which a pair of laser pulses can be used to prepare excited states that are not accessible with a single laser pulse.

## II. METHODS

The experiment was performed using a table-top keV-UED instrument, which has been described in detail previously [16, 46, 47]. In the standard configuration, a femtosecond laser pulse (the pump) triggers a photoinduced reaction, and a femtosecond electron pulse (the probe) is used to interrogate the molecules. As shown in Figure 1, we have modified the setup to accommodate a pair of time-delayed pump pulses. The diffracted electrons, which contain information about the changing molecular structure, are recorded by a two-dimensional detector. The electron pulse is generated by shining a femtosecond UV laser pulse onto a photocathode. The electron pulse is accelerated to 90 keV in a static electric field and then temporally compressed at the sample position using an RF cavity. Two important modifications in the setup were implemented to carry out this work. First, the temporal resolution was improved to 150 fs through the use of more stable synchronization electronics, and second, the optical setup was modified to accommodate two collinear pump laser pulses with a tilted pulse front at two different wavelengths. The tilted pulse front is necessary to avoid broadening of the temporal resolution as the laser and electron pulses traverse the sample with different speeds. Figure 1(a) shows a diagram of the interaction region including the incident electron and laser pulses, the gas jet and the scattered electrons which produce the diffraction patterns on the detector.

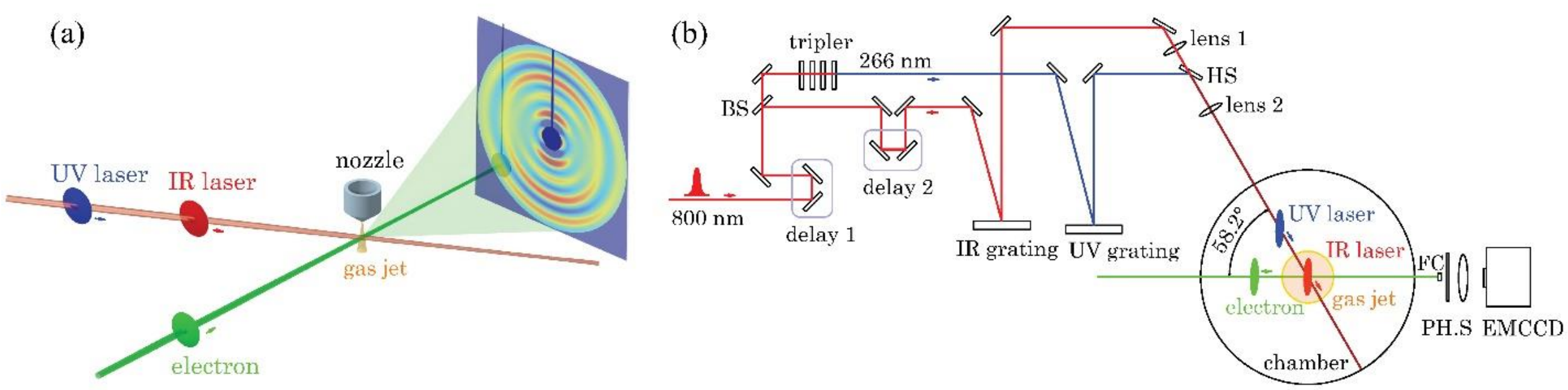

Figure 1. (a) Diagram of the UED experiment at the interaction region, including the gas jet, the electron pulse (green), the IR laser pulse (red) and the UV laser pulse (blue). (b) Schematic of the experimental layout with IR+UV laser pulses (not to scale). The path drawn in red represents the 800 nm-wavelength beam, while blue corresponds to the 266 nm laser beam, and green corresponds to the electron beam. The electron generation and pulse compression are not shown in this diagram, refer to APPENDIX A or [16] for details. A sketch of the electron, IR and UV beams shows how the velocity mismatch between electron and laser pulses is compensated, in which red, blue and green represent the IR, UV and electron pulses, respectively. Arrows indicate the propagation directions. BS = beam splitter, HS = harmonic separator, FC = Faraday cup, PH.S = phosphor screen.

Figure 1(b) shows a schematic diagram of the optical beam path used to deliver two collinear time-delayed pulses on the sample. An IR laser pulse (800 nm, 7 mJ, 50 fs FWHM) is produced by a Ti: Sapphire laser at a repetition rate of 1 kHz. Each laser pulse is split into three separate beams to provide the IR pre-excitation, to generate the UV for photodissociation, and to generate a second UV beam at 266 nm to trigger electron emission from the photocathode in the electron gun (not shown in the figure). The two UV pulses are generated through frequency tripling of the 800 nm beam with BBO crystals. The group velocity mismatch between laser and electron pulses [48, 49] is compensated by introducing an angle between the lasers and electron beams and by tilting the pulse intensity front of the IR and UV laser pulses with respect to their phase fronts [16, 50]. The tilted front is generated by reflecting the laser pulses from a diffraction grating and imaging the grating surface onto the sample plane, as shown in Figure 1(b). A more detailed description of the optical setup is presented in APPENDIX A.

The IR laser pulses on the sample have an energy of 2 mJ per pulse, a pulse duration of ~100 fs FWHM, and spot size of 240 μm FWHM, which results in a peak intensity of approximately $1.5\times10^{13}$ W/cm$^2$. The UV pulse has an energy of 100 μJ, a pulse duration of 120 fs FWHM, and spot size of 200 μm FWHM. The UV and IR pulses are both polarized in the same direction, which is perpendicular to the plane defined by the propagation directions of the electron and laser beams. We use delay 2 in Figure 1(b) to adjust the relative time delay between the IR and UV laser pulses. Once this time delay is optimized delay 2 is fixed, and delay 1 is varied to change relative time delay between electron pulse and pair of laser pulses.

The electron beam is guided by a pair of magnetic lenses and deflectors and is truncated by a 300 μm platinum aperture to deliver an electron beam current of 8 pA on the sample, corresponding to 50,000 electrons per pulse. The electron beam diameter is 200 μm FWHM on the sample and the pulse duration is estimated to be 100 fs or less. The instrument response function has been characterized to be 140 ± 10 fs (see details in APPENDIX C). The gas sample, $CF_3I$, has a purity of 99% and was purchased from Sigma-Aldrich. A de Laval nozzle with inner diameter of 30 μm is used to deliver the sample into the chamber with a backing pressure of 320 torr. The size of the gas jet at the interaction region is estimated to be 250 μm FWHM. The gas jet is orthogonal to the plane defined by the propagation directions of the laser and electron beams. The scattered electrons are captured by a phosphor screen that is imaged onto an electron-multiplying charge-coupled device (EMCCD). The directly transmitted electrons are stopped by a beam stop (Faraday cup) that measures the current of the transmitted electron beam.

### III. 3. ELECTRON SCATTERING THEORY

Here we give a brief overview of the theory of electron scattering and the modified pair distribution function (MPDF) that are relevant for this work (more details can be found APPENDIX B). In electron scattering from molecules in the gas phase, electron waves scattered from atoms within the molecule interfere, and the total scattering intensity is an incoherent sum of the diffraction signals from all the molecules in the ensemble. The total scattering intensity $I_{total}$ for a sample of molecules with an anisotropic angular distribution is given by [16, 51]:

$$I_{total}(\boldsymbol{s}) = \iint \sum_{j=1}^{n} \sum_{k=1}^{n} f_j^*(s) f_k(s)\, e^{-i\boldsymbol{s}\cdot\boldsymbol{r}_{\mathrm{jk}}(\alpha,\beta)} \times g_{jk}(\alpha)\, \sin\alpha\, \mathrm{d}\alpha \mathrm{d}\beta, \quad (1)$$

where $f_k(s)$ is the scattering amplitude of the $k^{\text{th}}$ atom within the molecule, $n$ is the total number of atoms that constitute the molecule, $\boldsymbol{r}_{jk}$ is a vector pointing from the $k^{\text{th}}$ atom to the $j^{\text{th}}$ atom with length equal to the interatomic distance, $g_{jk}(\alpha)$ is the angular distribution of atom pair $jk$, and $\boldsymbol{s}$ is the momentum transfer vector of the scattered electron. The atomic scattering intensity $I_A(s)$, which is independent of the molecular structure, is given by the sum of the terms with $j = k$, whereas molecular scattering intensity $I_M(\boldsymbol{s})$ is the sum of terms with $j \neq k$, in which the structural information and angular distribution are encoded. We isolate the time-dependent signal through the diffraction-difference method, given by the difference molecular scattering $\Delta I_M$

$$\Delta I_M(\boldsymbol{s}, t) = I_{total}(\boldsymbol{s}, t) - I_{total}(\boldsymbol{s}, t_0), \quad (2)$$

where the variable $t$ denotes the time delay with respect to the laser excitation, $I_{total}(\boldsymbol{s}, t_0)$ is the total scattering intensity at time $t_0$,which corresponds to a time before the arrival of the laser pulse, i.e. before the sample is excited [17, 52]. To separate the changes in molecular structure and angular distribution, we decompose the diffraction pattern into Legendre polynomials [53],

$$\Delta I_M(\boldsymbol{s}, t) = \Delta I_{M,0}(s, t) + P_2(\cos\phi)\Delta I_{M,2}(s, t), \quad (3)$$

where $s$ is the magnitude of the momentum transfer vector $\boldsymbol{s}$, $\phi$ is the azimuthal angle on the detector plane, and $P_2(\cos\phi)$ is the 2nd order Legendre polynomial. $\Delta I_{M,0}(s,t)$ and $\Delta I_{M,2}(s,t)$ are the isotropic and anisotropic components of the molecular scattering, respectively, with their analytic forms given in a prior publication [53]. Here $\Delta I_{M,0}(s,t)$ captures only the structural changes, while $\Delta I_{M,2}(s,t)$ captures changes in both structure and angular distribution. The contribution from higher order Legendre polynomials is zero for one-photon excitation, and is negligible for the degree of alignment generated by the IR pulse in our experiment. For the double pulse excitation, higher orders in the Legendre projection are in principle present but were not used in the reconstruction because their amplitude was found to be very small compared to the lower orders. The Legendre polynomial decomposition can also be used to filter out noise and artifacts in the diffraction signal [17, 54]. The normalized difference signal is defined as $\mathcal{R}_m(s,t) = \Delta I_{M,m}(s,t)/I_0(s)$ , where $m = 0,2$ indicates 0th or 2nd order terms, respectively, and $I_0(s)$ is the azimuthally averaged $I_{total}(\boldsymbol{s}, t_0)$.

The MPDF, which contains both structural and angular information in real space, is produced by the inverse Fourier transform, followed by the Abel inversion, of the rescaled molecular scattering intensity, formulated as $\mathcal{M}(\boldsymbol{s},t) = s^{\kappa} I_M(\boldsymbol{s},t)/I_A$, [3, 16, 51]

$$\mathrm{MPDF}(r,\alpha,t) = \mathrm{Abel}^{-1}\mathrm{FT}_{2\mathrm{D}}^{-1}\left[\frac{s^{\kappa} I_M(s,t)}{I_A}\right], \qquad (4)$$

where the value of $\kappa$ is adjusted to minimize artifacts in the transform and $\kappa$=0.01 is used below. The MPDF can be understood as an angularly dependent pair distribution function. The atom-pair distance and angular distribution of all atom pairs in real space can be retrieved from the MPDF (see APPENDIX B for details on the calculations of the MPDF). Furthermore, changes in the structure and angular distribution upon photoexcitation can be retrieved using the difference MPDF, $\Delta\mathrm{MPDF}(r,\alpha,t) = \mathrm{Abel}^{-1}\mathrm{FT}_{2\mathrm{D}}^{-1}[\Delta\mathcal{M}(\boldsymbol{s},t)]$, where $\Delta\mathcal{M}(\boldsymbol{s},t) = s^{\kappa}\Delta I_M(\boldsymbol{s},t)/I_A$.

## IV. RESULTS

We have carried out three sets of experiments to characterize the molecular response: IR excitation only, UV excitation only, and the combination of IR+UV excitation. We first describe the response of the molecule to the IR pre-excitation, then compare the dissociation dynamics after UV excitation with and without the IR pre-excitation.

### A. IR pre-excitation

We use a non-resonant femtosecond IR laser pulse which creates a ro-vibrational excitation in the electronic ground state of the molecule. We see evidence of both vibrations and alignment in the diffraction signal after excitation, and we do not observe evidence of dissociation or ionization. The laser pulse produces impulsive alignment of the $CF_3I$ molecules through an induced dipole interaction [55-58]. The maximum degree of alignment is reached approximately 1 ps after the laser excitation, followed by a rotational dephasing on a similar timescale and periodic revivals at later times. The diffraction signal becomes anisotropic when the molecules are aligned, as shown in Figure 2(a). The IR pulse can also produce a vibrational excitation through non-resonant impulsive stimulated Raman excitation (ISRS) [59-61]. The periods of the alignment, $u_2(t)$. The experimental data is represented by blue squares, and the simulation by the red dashed line. The right axis shows the corresponding values of $\langle\cos^2\theta\rangle$. The uncertainties in all figures are the standard deviation of the measurements and are obtained by the bootstrapping method.

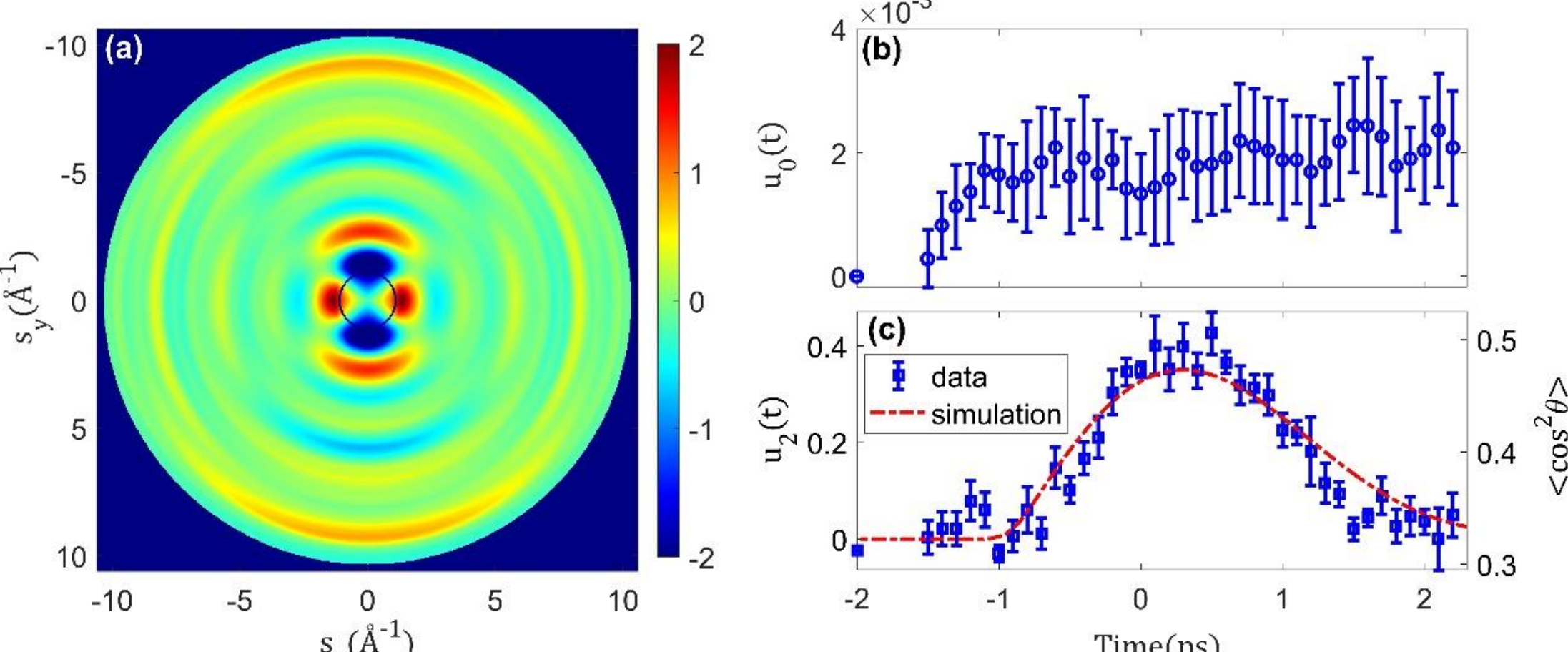

Figure 2. (a) Experimental pattern $\Delta\mathcal{M}(\boldsymbol{s})$ of aligned $CF_3I$ molecules at peak alignment, reconstructed from $\Delta I_{M,0}(s)$ and $\Delta I_{M,2}(s)$. The data at small s inside the black circle is not available experimentally and was filled in by extrapolation to 0 at $s$=0. (b) Time-dependent signal $u_0(t)$. (c) Time-dependent signal of nonadiabatic molecular

vibrational modes of $CF_3I$ are shorter than the instrument response function (IRF) of the UED instrument. Therefore, the motion cannot be directly time-resolved. However, the time-averaged diffraction signal of the vibrating molecule is different from the signal produced by a static molecule because it contains contributions from multiple structures.

A series of diffraction patterns $I_{total}(\boldsymbol{s}, t)$ with varying time delays between the IR laser and electron pulse, without the presence of the UV laser pulse, were recorded to capture the time evolution of the diffraction signal. The changes to the molecular structure and the angular distribution can be observed separately using a Legendre decomposition of $\Delta I_M(\boldsymbol{s}, t)$. We extracted the $\Delta I_{M,0}(s,t)$ and $\Delta I_{M,2}(s,t)$ from the measured diffraction patterns, as described by eqn (3) (details in APPENDIX D). Figure 2(a) shows the experimental $\Delta\mathcal{M}(\boldsymbol{s})$ of aligned $CF_3I$ molecules at the time-delay of maximum alignment. Here $\Delta I_M(\boldsymbol{s})$ was reconstructed from $\Delta I_{M,0}(s)$ and $\Delta I_{M,2}(s)$ using eqn (3). The main signature of alignment in the scattering pattern is the appearance of anisotropy, which contrasts against the concentric rings seen in the diffraction pattern of randomly oriented molecules. We use the amplitude of the isotropic and anisotropic parts of the signal to map the temporal evolution of the structural changes and the alignment, respectively.

Figure 2(b) shows the amplitude of the sum of the normalized difference signal $\mathcal{R}_0(s,t)$, integrated over a range of momentum transfer where the signal-to-noise ratio (SNR) is highest, from $s_{\min} = 2.3$ Å$^{-1}$ to $s_{\max} = 3.2$ Å$^{-1}$, denoted as $u_0(t) = \int_{s_{min}}^{s_{max}} \mathcal{R}_0(s,t)\, ds$. The signal amplitude rises over approximately 300 fs and saturates at a value corresponding to a 0.2% change in the diffraction intensity. The small amplitude of the signal indicates small structural changes are present, which could be due to small amplitude vibrations and/or minor changes in the equilibrium distances. The signal clearly indicates that there are changes taking place. The zero of time here is set as the time where the alignment signal (Figure 2(c)) starts to plateau. This setting is the most convenient for the later experiments that include both IR and UV laser pulses (see section IV.B).

Figure 2(c) shows the time evolution of the anisotropy, which closely tracks the degree of alignment defined as $\langle\cos^2\theta\rangle$, where $\theta$ is the angle between the laser polarization and the molecular axis. The integrated value of $\mathcal{R}_2(s,t)$, $u_2(t) = \int_{s_{min}}^{s_{max}} \mathcal{R}_2(s,t)\, ds$ over the range $s_{\min} = 2.3$ Å$^{-1}$ to $s_{\max} = 3.2$ Å$^{-1}$, was used to map the evolution of the alignment. We interpret the experimental signal by comparison with a theoretical calculation. We have calculated the evolution of the angular distribution of the molecules after the laser excitation by solving the Schrödinger equation (details in APPENDIX E). The angular distribution was then used to calculate the 2D diffraction signal, from which the $\mathcal{R}_2(s,t)$ was extracted using the same procedure as with the experimental data. The dashed red line in Figure 2(c) shows the results of the simulated $u_2(t)$, which is in very good agreement with the experimental data. Note that the left axis shows the time-dependent signal $u_2(t)$, and the right axis shows the corresponding values of the degree of alignment $\langle\cos^2\theta\rangle$. Figure 2(c) shows that the alignment takes approximately 1 ps to reach the maximum value, remains near the maximum for a few hundred femtoseconds and then decays. The isotropic signal $u_0(t)$ starts approximately 300 fs before the rise of the alignment, and it increases for about 300 fs before plateauing. This indicates that the structural changes are faster than the changes in the angular distribution, which agrees with vibrational motions being faster than rotation.

### B. Timing of UV excitation with respect to the IR pre-excitation

The first step before carrying out the two-pulse experiments is to set the relative time-delay between the UV and IR pulses. Recording the diffraction signal while the molecules are aligned has the advantage that the angular distribution can be used to distinguish between internuclear distances that are parallel and perpendicular to the alignment axis, which reveals more structural information [3]. Additionally, the excitation yield can be increased if the molecules are aligned such that the transition dipole moment is parallel to the polarization of the excitation pulse. For $CF_3I$ the transition dipole for excitation to the A-band points along the direction of the molecular axis (the C-I axis), which is the same as the alignment axis. One disadvantage of using alignment is that it is more challenging to retrieve information when the alignment is changing during the observation window. Thus, we select the time of arrival of the UV pulse ($t = 0$) to be the time where the alignment first reaches its maximum value. This provides us with a time window of approximately 500 fs after laser excitation to follow the reaction with a constant degree of alignment, which is sufficient to capture the dissociation dynamics. Note that the IR pulse arrives more than 1 ps before the UV pulse so there is no temporal overlap between the pulses, and the dissociation takes place in a field-free environment.

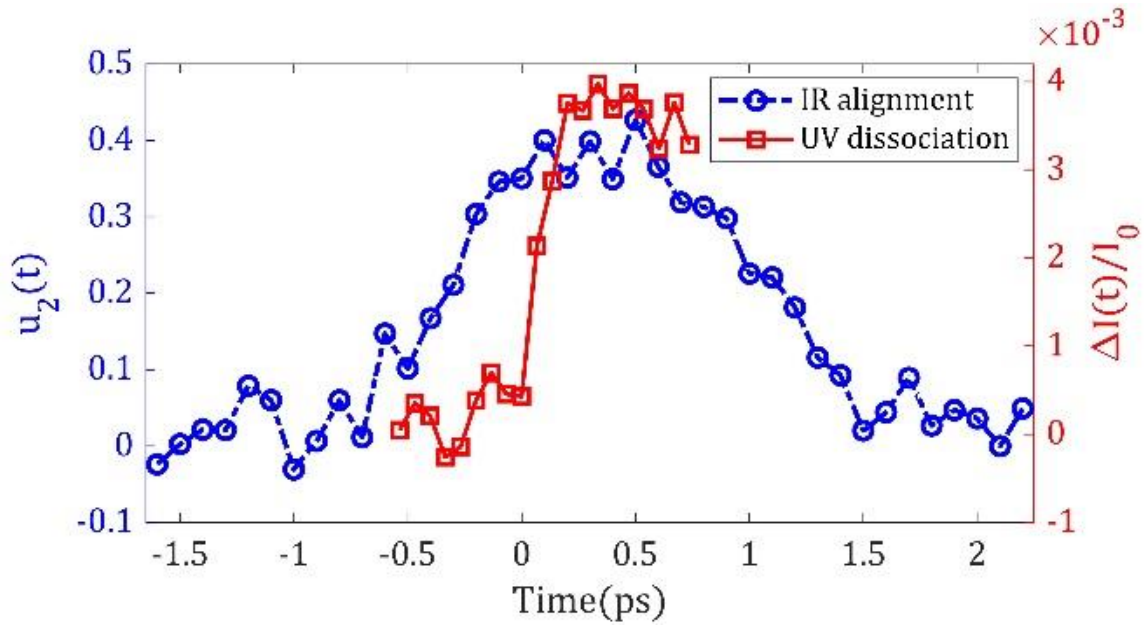

Figure 3. Synchronizing the maximum alignment in $u_2(t)$ (blue line) to the initial photodissociation characterized by the $\Delta I(t)/I_0$ (red line). Note that the left axis is for $u_2(t)$, and right axis for $\Delta I(t)/I_0$.

The time-delay between the IR and UV pulses was set by separately recording the dynamics after each pulse. As discussed earlier, the UV excitation leads to rapid photodissociation, which produces a clear signal in the diffraction pattern. We define the time zero for the UV-only signal as the time delay when the first changes in the $\Delta I(t)/I_0$ signal become visible. The evolution of the signal $\Delta I(t)/I_0$ is described in detail in APPENDIX C. We set the relative time delay between the IR and UV pulses such that the time zero of $\Delta I(t)/I_0$ induced by the UV pulse is synchronized to the beginning of the plateau in $u_2(t)$. Figure 3 shows the relative timing of the two signals, where the left axis corresponds to $u_2(t)$ and the right axis to $\Delta I(t)/I_0$. In the IR + UV experiment, the time delay between the two laser pulses is fixed while the time delay between the electron pulse and both laser pulses is varied using delay stage 1 in Figure 1(b).

### C. Photodissociation dynamics with IR pre-excitation

Here we compare the dissociation dynamics for the case of UV-only excitation with the case of dissociation for pre-excited molecules (IR+UV). In both cases, anisotropy is imprinted in the diffraction patterns because molecules along the laser polarization are preferentially excited. The transition dipole for UV excitation lies along the C-I bond direction, resulting in an angular distribution of the excited molecules that follows a $\cos^2\theta$ dependence, where $\theta$ is the angle between laser polarization and the C-I bond. Based on the amplitude of the diffraction signal, we estimate that 1.2% of the molecules are excited by the UV pulse. In the case of IR+UV excitation, the excited molecules are in a narrower angular distribution because there is a contribution from the pre-alignment of the IR pulse in addition to the angular selectivity of the UV excitation. We first look at the timescale of the photodissociation by extracting the $\mathcal{R}_0(s,t)$ using eqn (3). This signal captures the structural changes exclusively without the influence of the angular distribution. For the IR+UV measurement the $I_{total}(\boldsymbol{s},t_0)$ in eqn(2) used for the difference is recorded before the UV pulse but after the IR pulse, at a time ($t$ = -0.1 ps) when the molecules are already aligned, The $I_0(s)$ for the denominator is the same for both cases, recorded before both laser pulses. This is done to subtract the contribution to the signal from the alignment induced by the IR laser pulse for better comparison of the two cases.

Figure 4 (a)-(b) shows the time evolution of the normalized difference signal, $\mathcal{R}_0(s,t)$, for the UV-only and IR+UV excitation, respectively. The positive and negative features in the difference signals appear at the same locations in both cases. The signals have higher SNR at lower $s$ because the scattering intensity decreases rapidly as $s$ increases. The changes in the diffraction signal are the result of two structural changes: the separation of the two fragments after breaking of the C-I bond, and changes in the structure of the $CF_3$ fragment. The breaking of the C-I bond produces a significantly stronger signal than the changes in the structure of the fragment due to the comparatively high scattering cross section of the iodine atom.

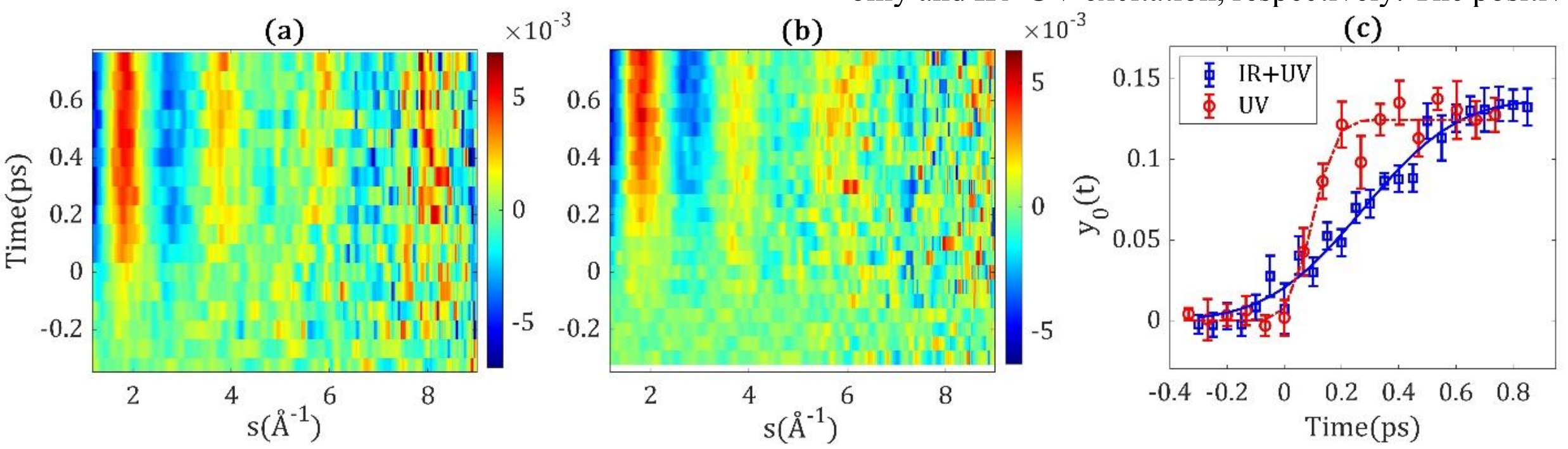

Figure 4. Normalized difference signal for UV-only and IR+UV excitation. (a) $\mathcal{R}_0(s,t)$ for $CF_3I$ photodissociation initiated in the UV-only experiment. (b) $\mathcal{R}_0(s,t)$ for IR+UV excitation. (c) Comparison of $y_0(t)$ for UV-only and IR+UV excitation. The red circles are data of the UV-only experiment, and the error function model with fitted parameters is the red dash line. The blue squares are the data of the IR+UV experiment, and the blue solid line is the error function with fitted parameters. The uncertainties are the standard deviation obtained by the bootstrapping method.

Figure 4 (a)-(b) shows similar signals in both cases because both excitations result in dissociation. There is, however, a clear difference in the timescale, with the UV-only signal having a significantly faster risetime. Figure 4(c) shows a lineout of the risetime of the diffraction signal, denoted as $y_0(t)$, which is calculated by integrating the $\mathcal{R}_0(s,t)$ over the two strongest peaks in the signal, specifically the regions between 1.3 $Å^{-1}$ to 2.0 $Å^{-1}$ and 3.1 $Å^{-1}$ to 4.3 $Å^{-1}$. Fitting an error function (APPENDIX C) to the $y_0(t)$ lineouts give a rise time of 150 ± 41 fs for the UV-only excitation, while for the IR + UV experiment the risetime is 618 ± 63 fs. The timescale of the UV-only data is comparable to the IRF of the UED setup, which suggests the dynamics are faster than 150 fs. In the second case, however, the dissociation is slowed down significantly by the pre-excitation. Based on the fits, we estimate an excitation percentage of 1.2% for the UV-only and 1.3% for the IR+UV cases. This is consistent with the expectation of a higher excitation yield with pre-aligned molecules. The slower reaction could be caused by a slower dissociation speed of the fragments or by a slower rate of dissociation, or a combination of both. We now investigate the signal in real space to extract more details about the process.

We use eqn (4) to generate the $\Delta\mathrm{MPDF}(r_x, r_y, t)$ for the UV-only and IR+UV measurements (more details on how the ΔMPDF was calculated can be found in APPENDIX F), where $r_x$ and $r_y$ are the components of the interatomic distances perpendicular and parallel to the alignment axis, respectively. The geometrical structure of $CF_3I$ in the ground state is shown in Figure 5(a). The experimental $\Delta\mathrm{MPDF}(r_x, r_y)$ at $t$ = 0.25 ps with IR+UV excitation is shown in Figure 5 (b), with blue indicating loss and red indicating gain of atom pair distances compared to the unexcited molecules. The polarization direction of both IR and UV is along the $r_y$ axis in the figure. The negative rings correspond to the loss of the ground state atom-pair distances: $r_{CF}$ = 1.33 Å, $r_{CI}$ = 2.14 Å, $r_{FF}$ = 2.15 Å, and $r_{FI}$ =2.89 Å, whereas the positive rings are due to the new atom-pair distances formed during the reaction. The atom-pair angular distribution is represented by the intensity distribution of the ring. The UV excitation of $CF_3I$ to the A-band is stronger for molecules with the C-I axis aligned along the polarization direction of the UV laser (more details in APPENDIX G). As a result, the signal from the C-I and F-I atom-pair distances, including both loss and gain, appear predominantly along the $r_y$ axis (corresponding to the azimuthal angle $\alpha$ = 0). The signal of the C-F and F-F atom-pairs appears along the $r_x$ axis, corresponding to the azimuthal angle $\alpha$ = π/2, perpendicular to the laser polarization. This angular separation is important because it separates the signals from the F-F and C-I distances, which overlap completely in the 1D-PDF.

The dissociation dynamics appear mostly along the direction parallel to the laser polarization, therefore we integrated the $\Delta\mathrm{MPDF}(r,\alpha,t)$ over the azimuthal angle $\alpha$ from 0° to 10° to capture the changes in the C-I and F-I distances using: $\Delta\mathrm{PDF}_{\parallel}(r,t) = \int_{0^\circ}^{10^\circ} \Delta\mathrm{MPDF}(r,\alpha,t)\, d\alpha$.

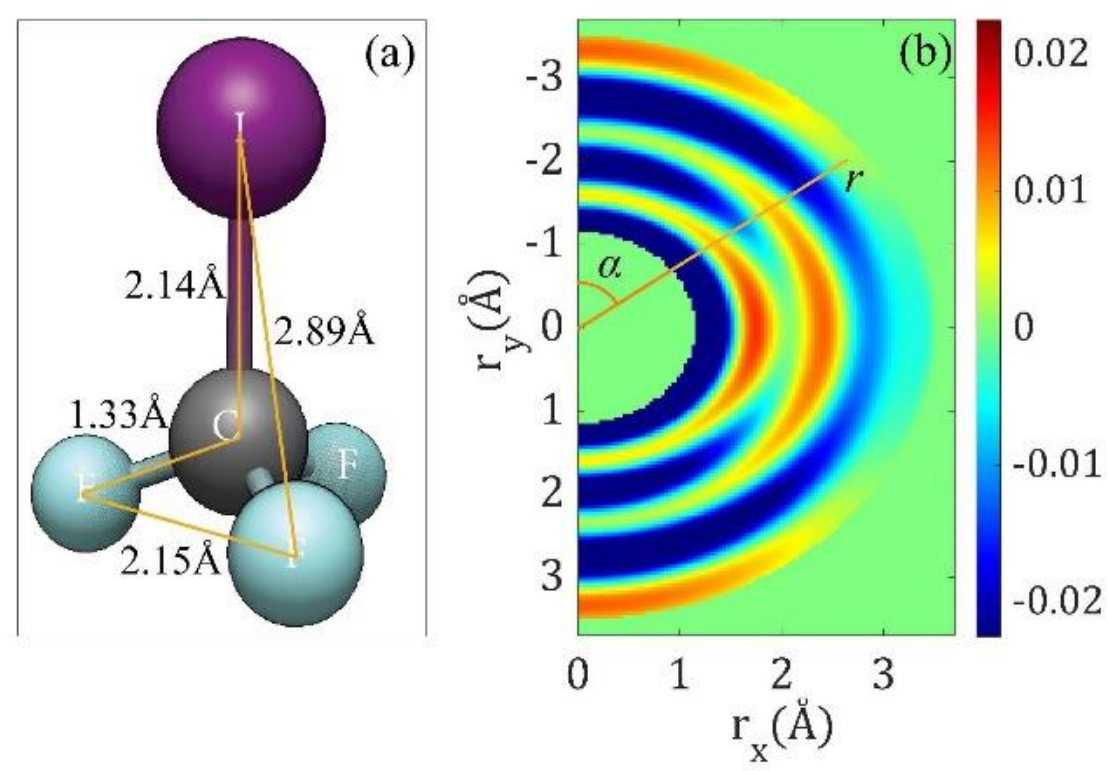

Figure 5. (a) Diagram showing the structure of $CF_3I$ in the ground state, with atomic labeling by the atom types and atom-pair distances $r_{CF}$ = 1.33Å, $r_{CI}$ = 2.14Å, $r_{FF}$ = 2.15Å, and $r_{FI}$ =2.89Å. (b) Experimental $\Delta\mathrm{MPDF}(r_x, r_y, t = 0.25\ \mathrm{ps})$ with IR+UV excitation. Blue indicates loss and red indicates gain of atom pair distances compared to the unexcited molecules. Polarization of IR and UV are along the $r_y$ axis, $r = \sqrt{r_x^2 + r_y^2}$ and $\alpha = \tan^{-1}(r_x/r_y)$.

Figure 6 (a) and (b) shows the $\Delta\mathrm{PDF}_{\parallel}(r,t)$ for the UV-only and IR+UV cases, respectively, over the time range from -0.1 ps to 0.4 ps. For the later times, both measurements show the depletion signals around $r_{CI}$ = 2.14 Å and $r_{FI}$ =2.89 Å indicating the loss of C-I and F-I distances due to the dissociation. However, the two signals are very different for the early times. The tilted bands that appear in Figure 6(b) are due to the gradual increase of the distance between the iodine and $CF_3$ fragment. The clearest evidence is the bright band which starts around 3 Å at $t$ = 0 and moves towards larger distances at later times. This corresponds to the dissociating wavepacket projected into the increasing F-I distance. The same tilted features are not observed in the case of the UV-only excitation (Figure 6(a)) because the wavepacket moves faster and the signal is blurred by the IRF. The dissociation with UV-only excitation is known to take place on a time scale of 50 fs [3, 10]. The fast dynamics are reflected in the appearance of the depletion features, and as shown in Figure 4(c). The depletion signals reach a plateau after 0.2 ps for UV-only excitation. The slower dissociation in IR+UV experiments is also evident in the slow increase in the amplitude of the depletion signals, which is consistent with the analysis shown previously in Figure 4(c). The real-space analysis, however, allows us to capture the motion of the dissociating wavepacket. This suggests that the dissociation process is still predominantly taking place coherently, with a slower dissociation speed of the fragments. The real-space analysis also shows that the direction of the transition dipole moment remains unchanged by the IR pre-excitation since the excited molecules show the same angular distribution peaked along the polarization of the UV pulse.

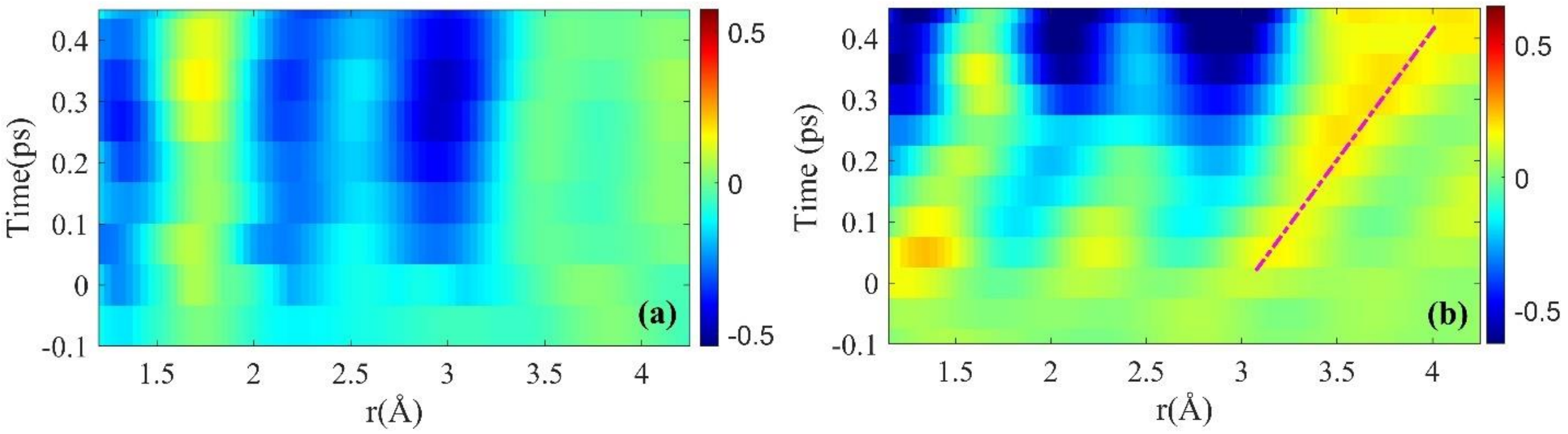

Figure 6. (a) Experimental $\Delta\text{PDF}_{\parallel}(r,t)$ for UV-only excitation, with blue indicating loss and red indicating gain of atom pair distances compared to the unexcited molecules. (b) Experimental $\Delta\text{PDF}_{\parallel}(r,t)$ for IR + UV excitation. The positive tilted distribution is the F-I wavepacket during the dissociation. The linear function with fitted parameters is the red dashed line.

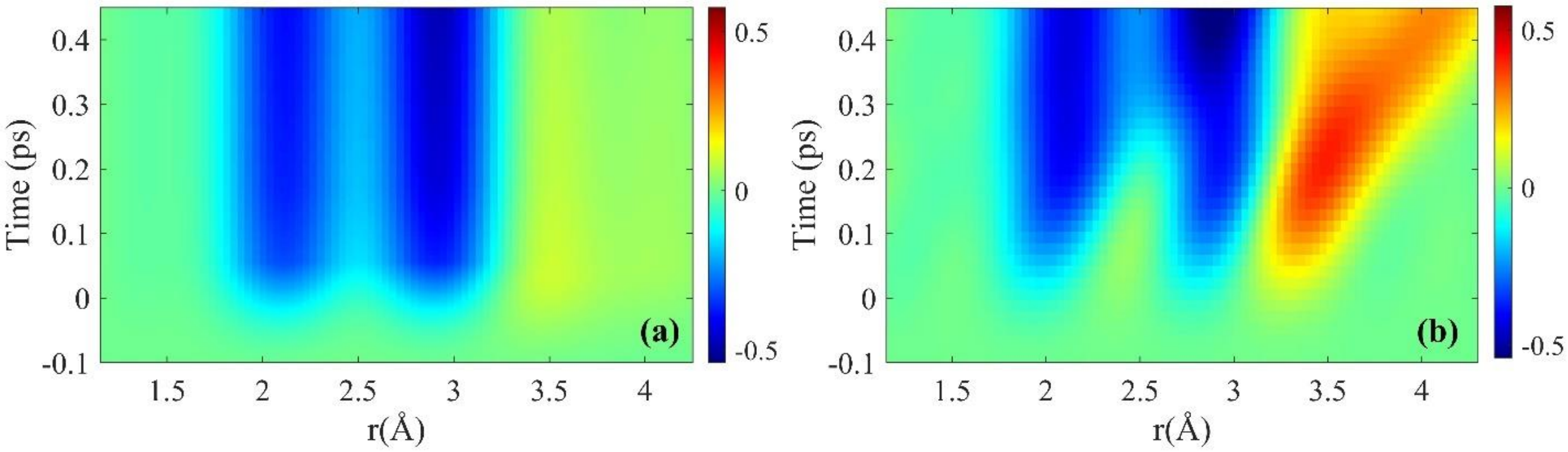

Figure 7. (a) Modeled $\Delta\text{PDF}_{\parallel}(\text{r},\text{t})$ convolved with a 0.15 ps FWHM instrument response function for the UV-only experiment. (b) Modeled $\Delta\text{PDF}_{\parallel}(\text{r},\text{t})$ convolved with a 0.15 ps FWHM instrument response function for the IR+UV excitation.

We extract a dissociation speed of 2.82 ± 0.57 Å/ps based on the movement of the peak corresponding to the F-I wavepacket, by a linear fit to the center position of the peak vs time. The red dashed line in Figure 6(b) shows the linear fit to the data. This speed is markedly slower than the dissociation speed for the UV-only excitation.

In order to validate our interpretation of the real-space signal, we use a simple model in which the iodine atom moves away along the direction of the C-I bond after UV excitation while the structure of the $CF_3$ fragment remains unchanged. While in reality we expect the $CF_3$ fragment to be vibrationally excited, our model focuses on capturing the effect of the dissociation on the diffraction signal. We first calculated $I_{total}(\boldsymbol{s},\Delta\text{CI})$, where $\Delta$CI represents the increase of the C-I atom-pair distance [62]. Applying the procedures used on the experimental data, we calculated $\Delta I_M(\boldsymbol{s},\Delta\text{CI})$ and $\Delta\text{MPDF}(r_x,r_y,\Delta\text{CI})$ to obtain the $\Delta PDF_{\parallel}(r, \Delta CI)$, as a function of $\Delta$CI (more details in APPENDIX H). For modelling the UV-only case, the relative speed between the fragments (dissociation speed) was set to 22.5 Å/ps, corresponding to 1.17eV kinetic energy release, based on previously published *ab-initio* calculations [63]. This dissociation speed is consistent with the previously reported translational energy of 1.15 eV for photofragments produced by UV induced photodissociation of $CF_3I$ [64, 65]. Using this dissociation speed, we converted the $\Delta$CI to time to obtain the modeled $\Delta PDF_{\parallel}(r, t)$. The final step is to apply a convolution in time to match the IRF of the measurement. For this we use a convolution with a Gaussian function with a FWHM of 150 fs. The same procedure was applied to model the IR+UV data, but in this case, we used the speed of 2.82 Å/ps which was retrieved from the experimental data.

Figure 7 (a) and (b) show the predicted $\Delta PDF_{\parallel}(r, t)$ for the UV-only and IR+UV cases, respectively, using the simple model. For the UV-only case the dissociation signal can be seen clearly before the convolution (not shown here, see APPENDIX H for the figure) but is completely blurred out when the IRF is considered (Figure 7(a)). The calculated signal agrees very well with the experimental signal in Figure 6(a), with the exception of the appearance of additional signals at shorter distance ($r < 1.5$ Å) in the experiments. These differences are caused by changes in the structure of the $CF_3$ fragment and also by artifacts that appear at short distances due to the finite *s*-range of the data (more information in APPENDIX H).

Figure 7 (b) shows the calculated signal corresponding to the slower dissociation speed of the IR+UV experiment. The main features in the experimental signal are reproduced by the model, with the tilted bands appearing at early times, and the strong band from the increasing F-I distance moving towards larger distances. This model validates our interpretation of a slow dissociation speed. While the model captures the dissociating wavepacket accurately, it does not capture all the details of the dynamics over the first 200 fs. These differences are likely due to vibrations and changes in the $CF_3$ structure, and possibly due to an initially non-uniform speed of the wavepacket. We have also tested a different model in which the dissociation speed is the same as in the UV-only case, but we assume a slower statistical dissociation taking place over several hundred femtoseconds, and this model qualitatively disagreed with the experimental data. Thus, we conclude that the dominant process is a coherent dissociation with slower moving fragments. The dissociation wavepacket observed corresponds to the F-I distance, which reaches a maximum value of approximately 4 Å during the measurement time window. This corresponds to an increase in the C-I bond distance of approximately 1 Å. Thus, it is possible that the wavepacket at this time is still moving in a shallow potential and that the final speed of the fragments will be larger. What is clear from this measurement is that the wavepacket is traversing a different region of the potential energy surface (PES).

The IR pre-excitation results in a new dissociation channel with slower speed, however, the data indicates that a fast dissociation channel might also be present. While we cannot directly see the wavepacket in the fast dissociation channel, we can still distinguish the two channels based on how fast the depletion features appear. This is clear in Figures 7 (a) and (b), where we assume only fast and slow dissociation, respectively. The depletion features in Figure 6 (a) quickly reach their maximum value while in Figure 6 (b) they continue to rise over several hundred femtoseconds. The experimental data for the UV-only excitation (Figure 6 (a)) agrees well with the calculation of the fast dissociation only in Figure 7 (a). The depletion signal in the data for the IR+UV excitation shown in Figure 6(b) is best matched by a combination of the fast and the slow channels. We estimate the contribution of the two channels by calculating the ratio of the amplitudes of the signal corresponding to the F-I dissociation wavepacket to the F-I depletion signal. The first originates only from the slow channels, while the second has contributions from both channels. We use the models shown in Figure 7 (b) to calibrate the signal and extract the relative yield from the experiments, which gives a yield of 40% ± 5% for the slow channel (more information in APPENDIX I). The presence of the fast channel could indicate that both channels are present after the IR+UV excitation, or it could originate from molecules that are only excited by the UV and not the IR pulse as a result of imperfect spatial overlap between the laser pulses on the sample.

## V. DISCUSSION

In the previous sections, we have clearly demonstrated the capability to carry out a GUED experiment with a pair of time-delayed pump laser pulses, and that the presence of the IR pre-excitation alters the reaction dynamics significantly. The IR pre-excitation opens a new dissociation channel through a region of the PES where the potential gradient is small compared to the A-band states, as evidenced by the lower dissociation speed.

We now discuss possible mechanisms by which the IR pre-excitation can alter the dynamics. We believe the most likely mechanism is that vibrational pre-excitation of the molecule through ISRS [60, 61, 66] results in the UV pump pulse populating a different excited state, or a different region of the PES where the slope of the potential is less steep and thus the fragments separate more slowly. The bandwidth of our femtosecond laser pulse is sufficient to drive ISRS in the lower energy vibrational modes of $CF_3I$. Previous experiments on vibrationally mediated photodissociation (VMP), have shown that pre-excitation can lead to new dissociation pathways and influence the energy of the fragments [67]. In these experiments a laser first prepares the molecules to a vibrationally excited state through resonance [68-74] or stimulated Raman excitation [59-61] in the electronic ground state, and subsequently a second laser transfer the population from the ground state to an electronic excited state that is dissociative. Interestingly, we have not found an example of VMP where the pre-excitation leads to a slower dissociation speed, although previous measurements have captured only the final energies and not the dissociation dynamics. A different excited state, involving different parts of the PES, could be reached due to a change in the molecular symmetry or the spread of the wavefunction caused by vibrational excitation. Given that during our measurement time window the C-I bond length is only extended only by ~ 1Å, what we observe may not represent the final speed of the fragments but the first part of the dissociation process where the wavepacket moves slowly through a shallow part of the PES. Further studies that capture the final energy of the fragments and *ab-initio* calculations will be needed to fully understand this new reaction path.

We have also considered two other possible processes that could affect the dissociation dynamics: enhanced two-photon absorption of the UV laser and ionization by the IR laser through multiphoton absorption. Previous studies have observed that under intense laser conditions, the $CF_3I$ molecule can be excited to a Rydberg state by absorbing two UV photons simultaneously, and that in this state the population is transiently trapped in a higher state before coupling to a dissociative state [3]. The transition dipole for this two-photon excitation, however, is perpendicular to that of the one-photon excitation to the dissociative A band, i.e. it is perpendicular to the molecular axis. Our analysis of the experimental ΔMPDF clearly shows that the dissociation takes place along the C-I axis. Therefore, it is not through the previously observed Rydberg state. In addition, we expect a low probability of multiphoton excitation because the UV laser intensity was three times lower than for the experiment of Yang *et al* where the two-photon excitation to the Rydberg was observed [3].

In the measurements with IR excitation only we observed no evidence of dissociation or ionization. Dissociation would have produced a clear signature in the diffraction signal, which was not present in the data. Ionization by an intense laser field produces macroscopic electric and magnetic fields that deflect and distort the electron beam [75-77], which we did not observe in the experiment. At the sample density of our experiments the deflection would have been strong and easily observable. While we cannot completely dismiss the probability of ionization, the fraction of ionized molecules would be too low to account for the different dissociation dynamics observed in the experiment.

## VI. CONCLUSIONS

This work extends the existing capacities of GUED to reach unexplored regions of the PES through double excitation, and to capture dynamics in the molecular frame through photoexcitation of aligned molecules. Measurements where a second laser pulse is used to alter the outcome of a reaction triggered by the first laser pulse are also possible, where the reaction would be tracked taking advantage of the real-space sensitivity of GUED. We developed the methodology to carry out experiments with two excitation laser pulses and to analyze and interpret the data, and to retrieve real space information in the molecular frame. A proof-of-principle experiment shows significantly different dynamics in the UV photodissociation of $CF_3I$ after pre-excitation with an IR femtosecond pulse. We take advantage of the real space and angular resolution of GUED to retrieve the structural changes and directionality of the dissociation. The results show that the IR pre-excitation opens a new dissociation channel that traverses a shallower region of the PES.

## ACKNOWLEDGMENTS

This work was supported by the US Department of Energy, Office of Science, Basic Energy Sciences under award no. DE-SC0014170.

## DATA AVAILABILITY

The data that support the findings of this article are available in [78].

## APPENDIX A: EXPERIMENTAL SETUP

Figure 8 shows a schematic diagram of UED experimental layout. A Ti: Sapphire laser generates laser pulses at a wavelength of 800 nm with 7 mJ pulse energy and 50 fs FWHM pulse duration at a repetition rate of 1 kHz. The laser output is split into three beams. Approximately 50% of the IR laser pulse energy is directed to the IR pre-excitation beam, 40% is used to generate a 266 nm pulse to initiate photodissociation of the sample molecules, and 10% to generate another 266 nm pulse for illuminating the photocathode to produce the electron pulses. The two 266 nm UV pulses are generated through frequency tripling of the IR pulses with BBO crystals. Group velocity mismatch between laser and electron pulses has been one major issue that limits keV-GUED to reach a femtosecond temporal resolution [48, 49]. In this experiment, the group velocity mismatch is compensated by introducing an angle between the laser and electron beams and by tilting the pulse intensity front of the IR and UV laser pulses with respect to their phase fronts [16, 50]. Both the tilted angle of laser pulses and cross angle between the lasers and electron beams are designed to be $\cos^{-1}(v_e/c)$, where $v_e$ and $c$ are speed of electron and laser pulse, respectively. The speed of electrons with a kinetic energy of 90 keV is 0.53 $c$, and the cross angle and tilted angle are 58.2°. The inset of Figure 8 shows a sketch of how the velocity mismatch between electron and laser pulses is compensated at the interaction region using a laser pulse with a tilted intensity front traveling at an angle of 58.2° with respect to the electron beam.

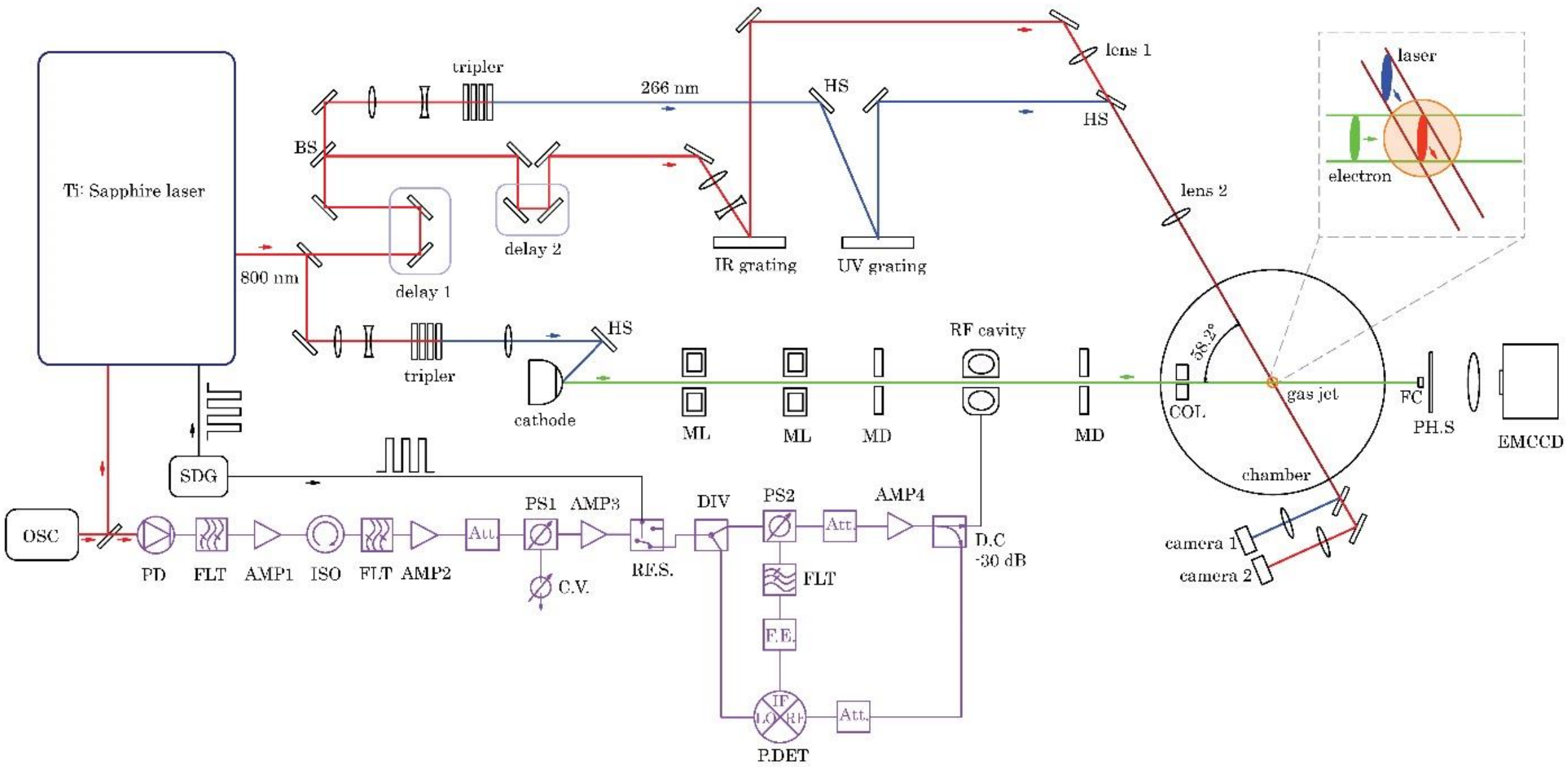

Figure 8. Diagram of UED experimental setup, not to scale. The path drawn in red, blue correspond to the 800 nm, 266 nm wavelength laser beam, respectively. The path drawn in green corresponds to the electron bunch trajectory from the photocathode to the detector. The path in black is the electric 1 kHz trigger signal from the laser control unit (SDG) to control the release timing of laser pulse and 3 GHz pulse. The path in purple corresponds to the 3 GHz RF generation and synchronization system used to drive the RF cavity. The inset shows a sketch of how the velocity mismatch is compensated at the interaction region using a laser pulse with a tilted intensity front travelling at an angle with respect to the electron beam. EMCCD captures the light generated by the diffracted electrons impinging on the phosphor screen. Cameras 1 and 2 are used to monitor the IR and UV laser beams. OSC = oscillator, SDG = signal delay generator, PD = photo diode, FLT = filter, AMP = amplifier, ISO = isolator, Att. = attenuator, RF.S = RF switch, DIV = power divider, PS = phase shifter, C.V.= control voltage, F.E.= feedback electronics, P.DET = phase detector, D.C. = directional coupler, ML = magnetic lens, MD = magnetic deflector, COL = collimator, FC = Faraday cup, PH.S = phosphorus screen, BS = beam splitter, HS = harmonic separator, EMCCD = electron-multiplying charge-coupled device.

To introduce the tilted angle in the UV laser pulse, we use an aluminum-coated grating with a grating constant of 400 mm$^{-1}$ and a lens (lens 2 in Figure 8 with a focal length 230 mm for 266 nm) that images the grating surface to the gas jet with a demagnification factor of 15.2. The incident angle of 6.1° allows the output UV beam to be perpendicular to the surface of grating [79]. We use a gold-coated grating with a constant of 150 mm$^{-1}$ and a two-lens system that images the grating surface to the gas jet with a demagnification factor of 13.5 to produce the tilted IR pulse. The output IR beam is normal to the surface of grating when the incident angle is 6.9°. The imaging system that consists of two lenses is required for the IR laser since the two lasers are collinear and the imaging relation for the two different wavelengths cannot be obtained using the same lens in the setup. The focal length of lens 1 and lens 2 are 2500 mm and 250 mm for 800 nm. The distance between lens 1 and lens 2 is around 150 mm to obtain an effective focal length of the two-lens system to be 235 mm to serve our purpose.

The electron generation and RF generation and synchronization system have been described in detail in [16]. In this experiment, two components in the RF generation and synchronization system reported in [16] have been updated to improve the phase stability of the 3 GHz time-varying electric field for electron pulse compression. Specifically, the low phase noise amplifier HMC8411LP2FE (AMP2 and AMP3 in Figure 8) and phase shifter HMC928LP5E (PS1) in [16] have been updated to Holzworth HX2400 and HX5100, respectively.

## APPENDIX B: ELECTRON DIFFRACTION THEORY

The total elastic scattering intensity from a single molecule can be written as,

$$I(\boldsymbol{s}) = \sum_{j=1}^{N} \sum_{k=1}^{n} f_j^*(s) f_k(s)\, e^{-i\boldsymbol{s}\cdot\boldsymbol{r}_{jk}}, \qquad \text{(B1)}$$

where $f_k(s)$ is the complex scattering amplitude of the $k^{\text{th}}$ atom, $N$ is the total number of atoms in the molecule, $\boldsymbol{s}$ is the momentum transfer vector with magnitude $s = \frac{4\pi}{\lambda}\sin\left(\frac{\theta}{2}\right)$, $\boldsymbol{r}_{jk}$ is the vector pointing from jth atom to kth atom. The terms in eqn (B1) where $j = k$ correspond to the atomic scattering intensity, formulated as $I_A(s) = \sum_{k=1}^{N} |f_k(s)|^2$, while the $j \neq k$ terms, called molecular scattering intensity, represent interference of atomic pairs. Note that $\boldsymbol{s}$ is momentum transfer vector for the 2D diffraction pattern, whereas $s$ is the magnitude of momentum transfer and is used for the 1D scattering intensity. The total scattering intensity $I_{total}$ for a sample of molecules with an anisotropic angular distribution is given by [16, 51]

$$I_{total}(\boldsymbol{s}) = \iint \sum_{j=1}^{n} \sum_{k=1}^{n} f_j^*(s) f_k(s)\, e^{-i\boldsymbol{s}\cdot\boldsymbol{r}_{\mathrm{jk}}(\alpha,\beta)} \times g_{jk}(\alpha,\beta) \sin\alpha \,\mathrm{d}\alpha \mathrm{d}\beta. \qquad \text{(B2)}$$

When the $g_{jk}(\alpha,\beta)$ only has dependence on $\alpha$, the diffraction intensity is given by eqn (1) in main text, i.e. in our experiment, linearly polarized lasers are used to pump the molecular ensemble. For a sample of molecules with randomly oriented distribution $g_{jk}(\alpha,\beta) = \frac{1}{4\pi}$, the 2D molecular scattering intensity $I_M(\boldsymbol{s})$ is given by [80]

$$I_M(\boldsymbol{s}) = \sum_{j=1}^{N} \sum_{k\neq j}^{N} f_j^*(s) f_k(s) \frac{\sin(s r_{jk})}{s r_{jk}}. \qquad \text{(B3)}$$

The molecular scattering intensity $I_M(\boldsymbol{s})$ can be azimuthally averaged to obtain the 1D scattering intensity $I_M(s)$ with a higher signal-to-noise ratio. Since scattering amplitudes $f_k(s)$ are approximately proportional to $s^{-2}$, the molecular scattering term $I_M(s)$ decays approximately as $s^{-5}$. Therefore, the modified scattering intensity is defined to enhance the oscillations in the signal:

$$sM(s) = \frac{sI_M(s)}{I_A(s)}. \qquad \text{(B4)}$$

In this work, the gas sample is pumped by linearly polarized laser pulses, and $I_M(\boldsymbol{s})$ does not decay approximately as $s^{-5}$ . For example, for perfectly aligned molecules given by eqn (B1), the $I_M(\boldsymbol{s})$ decays approximately as $s^{-4}$ . Therefore, the 2D modified scattering pattern of anisotropic molecular ensemble here is rescaled as

$$\mathcal{M}(\boldsymbol{s}) = \frac{s^{\kappa} I_M(\boldsymbol{s})}{I_A(s)}, \qquad \text{(B5)}$$

where $0 < \kappa < 1$ , and a higher anisotropy corresponds to a smaller $\kappa$. It has been shown in ref.[3, 16, 51] that the atom-pair distances and angular distribution of all atom pairs in real space can be retrieved from the modified pair distribution function (MPDF). The MPDF is generated by the Fourier inversion, followed by the Abel inversion, of $\mathcal{M}(\boldsymbol{s})$,

$$\mathrm{MPDF}(r,\alpha) = \mathrm{Abel}^{-1}\mathrm{FT}_{2\mathrm{D}}^{-1}[\mathcal{M}(\boldsymbol{s})] = \sum_{j=1}^{n} \sum_{k=1, j\neq k}^{n} g_{jk}(\alpha)\, H(r - r_{jk}) \otimes \frac{\tilde{F}_j(r) \star \tilde{F}_k(r)}{{r_{jk}}^2}, \qquad \text{(B6)}$$

where $0 < \kappa < 1$, and $\otimes$ signifies convolution and $\star$ stands for correlation; $\tilde{F}_j(r)$ is the Fourier transform of the normalized atomic scattering amplitude $f_j(s)/\sqrt{s^{-\kappa} I_A}$; $g_{jk}(\alpha)$ is the angular distribution of atom pair $\boldsymbol{r}_{jk}$; and $H(r - r_{jk})$ is the convolution of the PDF with the inverse Fourier transform of a function that truncates the diffraction signal to match the limited measurement range. The scattering pattern of a randomly orientated ensemble corresponds to $\kappa = 1$, where scattering pattern with a higher anisotropy corresponds to a smaller $\kappa$. The Fourier transform of the normalized atomic scattering amplitude $f_j(s)/\sqrt{s^{-\kappa} I_A}$ is much narrower than that of $f_j(s)$, which is helpful to isolate distances of different atom pairs.

In time-resolved experiments, we use the diffraction-difference method to enhance the change of diffraction pattern, formulated as $\Delta I_M(\boldsymbol{s},t) = I_{total}(\boldsymbol{s},t) - I_{total}(\boldsymbol{s},t_0)$, where $I_{total}(\boldsymbol{s},t_0)$ is the total scattering intensity at time delay $t_0$ recorded before the arrival of the laser pulse [17, 52]. The changes in molecular structure and angular distributions can be separated by decomposing $\Delta I_M(\boldsymbol{s},t)$ into Legendre polynomials, given by eqn (3) in main text. The normalized difference signal is defined as

$$\mathcal{R}_m(s,t) = \Delta I_{M,m}(s,t)/I_0(s), \qquad \text{(B7)}$$

where $m$ indicates 0th or 2nd order terms from the Legendre polynomial decomposition, and $I_0(s)$ is the azimuthally averaged $I_{total}(\boldsymbol{s},t_0)$. Correspondingly, we can define $\Delta\mathcal{M}(\boldsymbol{s},t) = s^{\kappa}\Delta I_M(\boldsymbol{s},t)/I_A$, and the ΔMPDF is given by

$$\Delta\text{MPDF}(r,\alpha,t) = \text{Abel}^{-1}\text{FT}_{2\text{D}}^{-1}[\Delta\mathcal{M}(\boldsymbol{s},t)]. \quad \text{(B8)}$$

We denote $p\left(r - r_{ij}\right) = H(r - r_{jk}) \otimes \frac{\tilde{F}_j(r) \star \tilde{F}_k(r)}{r_{\text{jk}}{}^2}$, and $\text{MPDF} = \sum_{i=1}^{N}\sum_{j=1,j\neq i}^{N} g_{ij}(\alpha)p\left(r - r_{ij}\right)$. The ΔMPDF is given by

$$\Delta\text{MPDF} = \sum_{i=1}^{N}\sum_{j=1,j\neq i}^{N} \tilde{g}_{ij}(\alpha)\Delta p\left(r - r_{ij}\right) + \Delta g_{ij}(\alpha)p\left(r - r_{ij}\right)$$

$$= \sum_{i=1}^{N}\sum_{j=1,j\neq i}^{N} \Delta g_{ij}(\alpha)\tilde{p}\left(r - r_{ij}\right) + g_{ij}(\alpha)\Delta p\left(r - r_{ij}\right) \quad \text{(B9)}$$

where $\tilde{p}\left(r - r_{ij}\right)$, $\tilde{g}_{ij}(\alpha)$ are the newly produced atom-pair distance and angular distribution of $\boldsymbol{r_{ij}}$ after the arrival of laser, whereas $g_{ij}(\alpha)$, $p\left(r - r_{ij}\right)$ are the terms before arrival of laser. The changes are given by $\Delta g_{ij}(\alpha) = \tilde{g}_{ij}(\alpha) - g_{ij}(\alpha)$ and $\Delta p\left(r - r_{ij}\right) = \tilde{p}\left(r - r_{ij}\right) - p\left(r - r_{ij}\right)$. The ΔMPDF contains direct information on the change of angular distribution $g_{ij}(\alpha)$ and atom-pair distance $p\left(r - r_{ij}\right)$. When the newly produced atom-pair distances can be isolated from other ones, the $g_{ij}(\alpha)$ can be retrieved from the ΔMPDF.

## APPENDIX C: INSTRUMENT RESPONSE FUNCTION

The instrument response time of the keV-GUED is characterized using the diffraction signal produced by the photodissociation of $CF_3I$ with a UV femtosecond laser pulse. The measured instrument response function (IRF) includes all contributions to the observed temporal resolution: electron and laser pulse durations, group velocity mismatch, timing jitter, and the intrinsic time of the dissociation process. The C-I bond cleavage in $CF_3I$ molecule upon absorption of one 266 nm photon happens within 50 fs, providing a good metric to estimate the overall temporal resolution of GUED [3, 10]. The photodissociation was captured in a series of electron diffraction patterns, denoted by $I_{total}(\boldsymbol{s},t)$, as a function of time delay between the UV laser and electron pulses. The time dependent $\Delta I(s,t)/I_0$ is generated with the following steps. First, a background image recorded with no sample is subtracted from all the diffraction images. Second, the 2D diffraction difference patterns $\Delta I(\boldsymbol{s},t) = I_{total}(\boldsymbol{s},t) - I_{total}(\boldsymbol{s},t_0)$ are calculated, where $t_0$ corresponds to a time before the arrival of the laser pulse, such that the sample is still in the ground state. Third, the $I_{total}(\boldsymbol{s},t_0)$ is average azimuthally to obtain $I_0(s)$. Fourth, the diffraction difference patterns $\Delta I(\boldsymbol{s},t)$ were divided into 4 quadrants, corresponding to two vertical and two horizontal cones with an open angle of 90º. This is done because the diffraction signal appears stronger along the direction of laser polarization due to the angular selectivity of the excitation. The diffraction data in the two vertical cones are azimuthally averaged to obtain the $\Delta I(s,t)$. The amplitude of $\Delta I(s,t)/I_0$ calculated from the vertical cones is 2 times larger than that of horizontal cones due to the anisotropy in the diffraction patterns.

The $\Delta I(s,t)/I_0$ is shown in Figure 9(a). The $\Delta I(s,t)/I_0$ was integrated over the range 1.3 Å$^{-1}$ to 2.1 Å$^{-1}$ to obtain time-dependent $\Delta I(t)/I_0$ shown in Figure 9(b). This function represents the risetime of the dissociation signal using the strongest feature in the diffraction pattern. The experimental $\Delta I(t)/I_0$, shown as the blue circles and the uncertainty was obtained using bootstrapping. An error function is used to fit the data and extract a time scale:

$$h(t) = \frac{c}{2}\left\{1 + \text{erf}\left[\frac{2\sqrt{\ln 2}(t - t_0)}{\tau}\right]\right\} \quad \text{(C1)}$$

where erf $(t)$ is the Gauss error function, $\tau, c, t_0$ are the parameters to fit, and $\tau$ characterizes the FWHM temporal resolution of the measurement, which we define as the IRF. The fit returns a value of $\tau = 140 \pm 10$ fs using bootstrapping method, which corresponds to the upper limit of the IRF. The retrieved $h(t)$ is shown by the red curve in Figure 9(b).

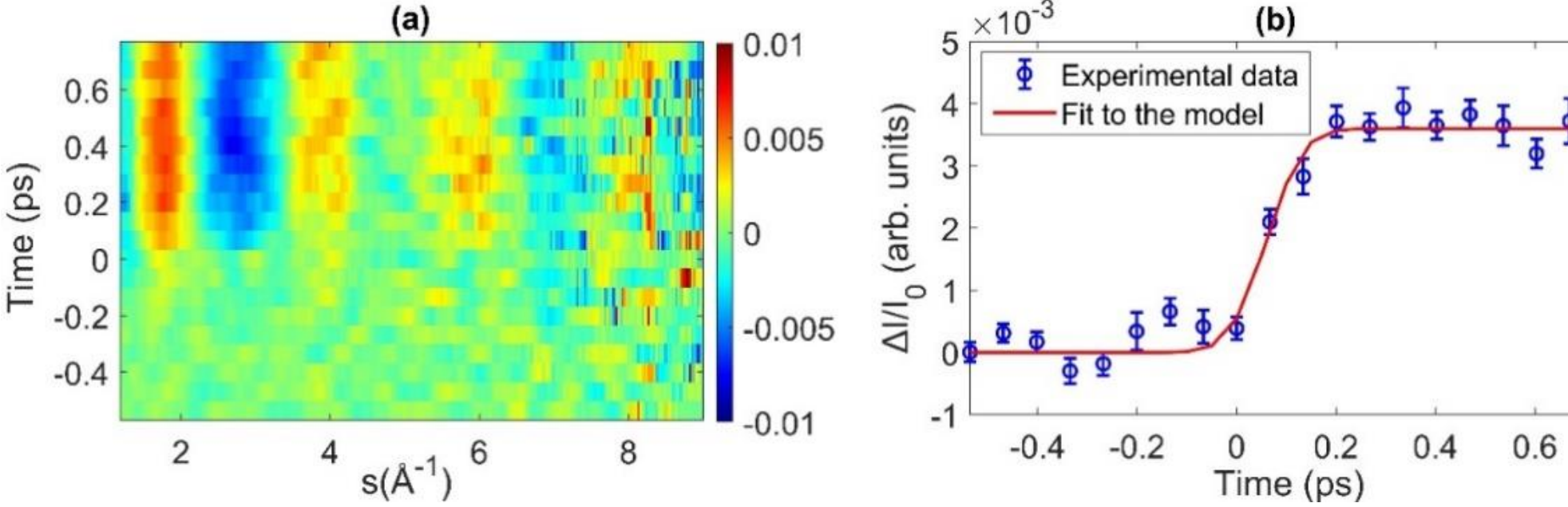

Figure 9. Temporal resolution characterization using the UV induced photodissociation dynamics of $CF_3I$ molecules. (a) Fractional difference signal $\Delta I(s,t)/I_0$. (b) The data from 1.3 Å$^{-1}$ to 2.1 Å$^{-1}$ is integrated to calculate $\Delta I(t)/I_0$ to characterize the instrument response time. The fitting of $\Delta I(t)/I_0$ to the error function model. The experimental data

is blue circles, and $h(t)$ with fitted parameters is the red curve. Statistical error bars are generated by bootstrapping at each data point.

## APPENDIX D: NORMALIZED DIFFERENCE SIGNAL

The normalized difference signals for IR-only, UV-only and IR+UV experiments are calculated with the following steps.

1. A background image recorded without sample is subtracted from each diffraction image.

2. The diffraction difference patterns $\Delta I_M(\boldsymbol{s},t)$ are calculated by taking the difference between $I_{total}(\boldsymbol{s},t)$ and $I_{total}(\boldsymbol{s},t_0)$ using eqn (2) in main text, where the variable $t$ denotes the time delay with respect to the laser excitation. For the IR-only and UV-only experiments, the $t_0$ corresponds to a time delay before the arrival of the pump laser, i.e. before the sample is excited. For the IR+UV experiment, both $I_{total}(\boldsymbol{s},t)$ and $I_{total}(\boldsymbol{s},t_0)$ are taken at the peak alignment of $CF_3I$ molecules produced by the IR laser, while the $I_{total}(\boldsymbol{s},t_0)$ is taken before the arrival of the UV laser. A diffraction pattern before the arrival of both IR and UV lasers was taken to calculate $I_0(s)$ for the denominator of normalized difference signal, given by eqn (B7), in IR+UV experiment.

3. The $\Delta I_M(\boldsymbol{s},t)$ are decomposed into Legendre polynomials using eqn (3) in main text.

4. The residual background in $\Delta I_{M,0}(s,t)$ is removed. A third order polynomial is first fit to the $\Delta I_{M,0}(s,t)$ to approximate the background, and then removed from the signal.

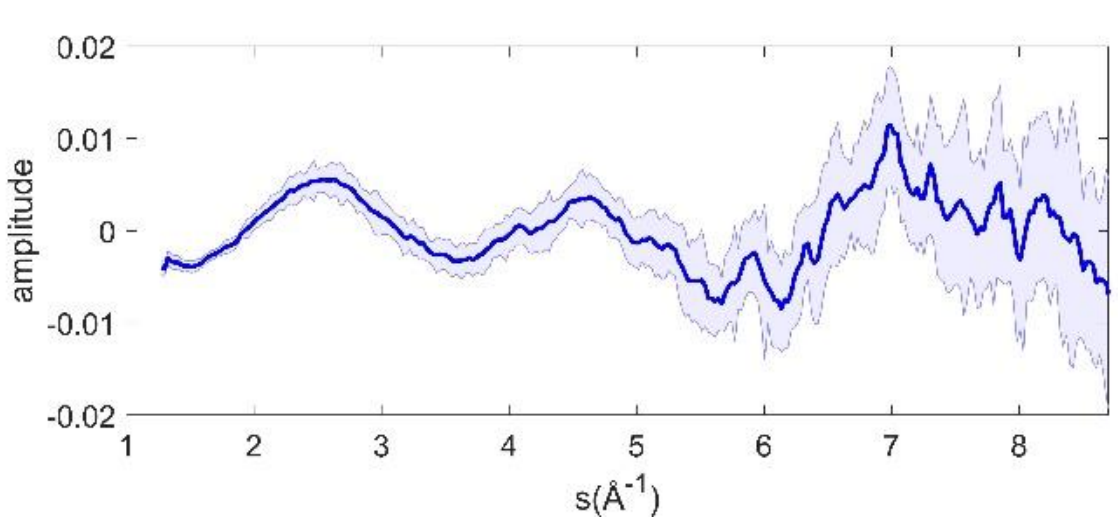

Figure 10. Normalized difference signal that indicates structure change of $CF_3I$ due to the IR excitation.

The IR-only excitation produces a very small signal, so in order to extract the amplitude of the signal we averaged the $\mathcal{R}_0(s,t)$ with time delay from -1 ps to 2.2 ps with a time step of 0.1 ps since the signal plateaus after -1 ps shown in Figure 2(b). We then performed a 35-point moving average in $s$ to remove high frequency noise. This produces a time-averaged signal with an amplitude of approximately 0.2%. The signal amplitude is too small to retrieve the signal as a function of time, i.e. at each time step individually. The time-averaged $\mathcal{R}_0(s,t)$ is shown in Figure 10. The blue shaded area in the figure corresponds to the standard deviation of the measurements obtained by the bootstrapping method.

## APPENDIX E: IMPULSIVE ALIGNMENT OF $CF_3I$ MOLECULES

Impulsive alignment of symmetric top molecules by a non-resonant, linearly polarized femtosecond laser is described by the time-dependent Schrödinger equation [56]

$$\mathrm{i}\hbar\frac{\partial\Psi(t)}{\partial t}=H(t)\Psi(t), \qquad \text{(E1)}$$

$$H_{\mathrm{ind}}=-\frac{1}{4}\mathcal{E}^2(t)\Delta p\cos^2\theta,\ \ \Delta \mathrm{p}=p_{\parallel}-p_{\perp}, \qquad \text{(E2)}$$

where $H(t)=H_{\mathrm{rot}}+H_{\mathrm{ind}}$, and $H_{\mathrm{rot}}=C_e\boldsymbol{J}^2+(A_e-C_e)J_z^2$, $H_{\mathrm{ind}}$ is the potential energy of the induced dipole in the electric field of the laser; the rotational constants $A_e>C_e$ are for a prolate symmetric top molecule, $\mathcal{E}(t)$ is the envelope of the electric field; $p_{\parallel}$ and $p_{\perp}$ are the polarizability tensor components that are parallel and perpendicular to the molecular axis, repectively; $\theta$ is the angle between molecular axis and the laser polarization direction, Equation (E1) can be solved numerically using the symmetric top eigenfunctions$|JKM\rangle$ as basis functions [81], with

$$|JKM\rangle=\left(\frac{2J+1}{8\pi^2}\right)^{\frac{1}{2}}e^{iM\phi}d_{MK}^{J}(\theta)e^{iK\chi}, \qquad \text{(E3)}$$

where $d_{MK}^{J}(\theta)$ is the Wigner's d-matrix, and $(\phi,\theta,\chi)$ are the Euler angles. The interaction of a linearly polarized laser field with a symmetric top molecule does not alter the quantum numbers $K$ and $M$ [56]. Therefore, the wave packet can be expanded as $\Psi_{J_iK_iM_i}(t)=\sum_J c^{JK_iM_i}(t)\,|JK_iM_i\rangle$ , where $|J_iK_iM_i\rangle$ are the intial states in an ensemble gorverned by the Boltzmann distribution $W_{J_iK_iM_i}(T)$ with a nuclear spin stastics weight [82], and $T$ is the temperature of the ensemble. The coefficients $c^{JK_iM_i}(t)$ are calcualted numerically. The degree of alignment, which is commonly used to describe the evolution of the alignment, is defined as

$$\langle\cos^2\theta\rangle(t)=\sum_{J_i,K_i,M_i}W_{J_iK_iM_i}(T)$$
$$\times\left\langle\Psi_{J_iK_iM_i}(t)\middle|\cos^2\theta\middle|\Psi_{J_iK_iM_i}(t)\right\rangle \qquad \text{(E4)}$$

The probability density of the aligned molecules is given by $\rho=\sum_{J_i,K_i,M_i}W_{J_iK_iM_i}(T)\left|\Psi_{J_iK_iM_i}(t)\right|^2$. In the case of symmetric top molecules aligned by a linearly polarized laser field, the probability density can be written as $\rho(\phi,\theta,\chi,t)=\left(\frac{1}{2\pi}\right)^2\rho_1(\theta,t)$ since it has

no dependence on $\phi$ and $\chi$. With the geometrical structure of the molecule and $\rho_1(\theta, t)$, the scattering pattern $I_M(\boldsymbol{s}, t)$ can be calculated with the method described in [83]. The time-dependent anisotropy signal $u_2(t)$ shown in Figure 2(c) can be calculated using $u_2(t) = \int_{2.3}^{3.2} \mathcal{R}_2(s, t)\ ds$. The temperature of the sample is not known accurately a priori, so it is determined by comparing the experimental and theoretical $u_2(t)$ using the fitting method described in [51].

## APPENDIX F: EXPERIMENTAL ΔMPDF

In this section we describe the procedure to retrieve the experimental ΔMPDF..

1.We start from the $\Delta I_M(\boldsymbol{s}, t) = I_{total}(\boldsymbol{s}, t) - I_{total}(\boldsymbol{s}, t_0)$ as described in APPENDIX D.

2.The $\Delta I_M(\boldsymbol{s}, t)$ are projected onto Legendre polynomials using eqn (3) in the main text.

3.The residual background in the $\Delta I_M(\boldsymbol{s}, t)$ is corrected using a 3rd order polynomial that is first fit and then subtracted from the $\Delta I_M(\boldsymbol{s}, t)$.

4. The region of missing data at low s (due to the beam stop) is smoothly interpolated to zero at the center of the pattern from the surrounding values [17, 28].

5.The $\Delta\mathcal{M}(\boldsymbol{s})$ is calculated using equation (B5) and a Gaussian damping term $e^{-0.025\times s^2}$ is applied to suppress the noise and artifacts at higher $s$ values where the SNR is lower [53].

6. The ΔMPDF is calculated using equation (4) in the main text using a momentum transfer range up to $s$ = 9.2Å$^{-1}$ and the zero-padding technique is applied in $\Delta\mathcal{M}(\boldsymbol{s})$ to decrease the step size of the ΔMPDF.

The calculation of the ΔMPDF and the $\Delta\text{PDF}_{\parallel}$ are outlined graphically in Figure 11. The $\Delta\mathcal{M}(\boldsymbol{s})$ in the IR+UV experiment at $t = 0.4$ ps is shown in Figure 11 (a). The polarization of both IR and UV lasers is along the y axis in Figure 11 (a). An obvious anisotropy imprints in the diffraction pattern. The data inside of the black circle ($s$ < 1.17Å$^{-1}$) is not accessible experimentally due to the presence of the beam stop that stops the directly transmitted electron beam. The missing data is filling in by extrapolating the data outside the region to zero at the center of the pattern.

The experimental ΔMPDF for $t$ = 0.4 ps is shown in Figure 11(b). The dashed lines correspond to the interatomic distances of the ground state $CF_3I$: $r_{\text{CF}}$ = 1.33 Å, $r_{\text{CI}}$ = 2.14 Å, $r_{\text{FF}}$ = 2.15 Å, and $r_{\text{FI}}$ =2.89 Å. The depletion signal of ground state F-I and C-I distances show a stronger signal along the vertical direction due to the selectivity of the photoexcitation where the excitation probability is highest for molecules aligned along the direction of laser polarization. The projection of the dissociation wavepacket onto the F-I distance can be clearly identified by the positive signal around 4 Å at $t = 0.4$ ps. The signal of F-F distance, described by the positive region around $r_{\text{FF}}$ = 2.15 Å that appears preferentially perpendicular to the y axis, indicates the hot vibrational motion of the $CF_3$ fragment. The signal of C-F is subjected to artifacts due to the interpolation of the missing data and residual background correction in the experimental pattern, which will be discussed in APPENDIX H. The ΔMPDF of aligned molecules provides information about the structural dynamics in the molecular frame.

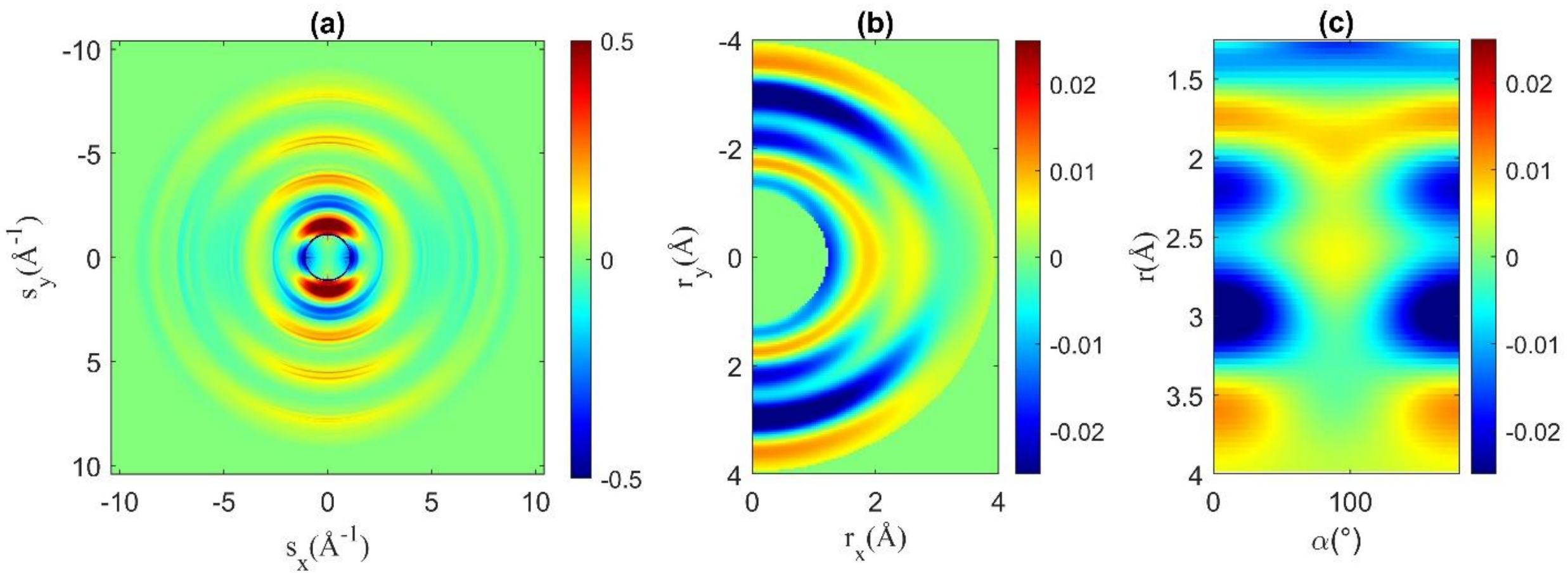

Figure 11. (a) $\Delta\mathcal{M}(\boldsymbol{s})$ for the IR+UV excitation. The data inside of the black circle, corresponding to the missing data due to beam stop, are interpolated smoothly to zero at the center of the pattern. (b) $\Delta\text{MPDF}(r_x, r_y)$. (c) Polar representation of $\Delta\text{MPDF}(r, \alpha)$.

Figure 11(c) shows the ΔMPDF in polar representation, in which both the internuclear distances and angular distribution of atom pairs can be clearly identified and be extracted. We focus on the structural changes along the polarization direction by integrating the signal over an angular range from 0° to 10° to generate the $\Delta\text{PDF}_{\parallel}$.

This function captures the signal of the C-I bond elongation dynamics primarily in the F-I interference signal which has a stronger signal than the C-I (there are three F-I distances that contribute to the signal and only one C-I distance). The $\Delta\mathrm{PDF}_{\parallel}(\mathrm{r},\mathrm{t})$ shown in Figure 6 in the main text was calculated over the range of -0.1 ps to 0.45 ps.

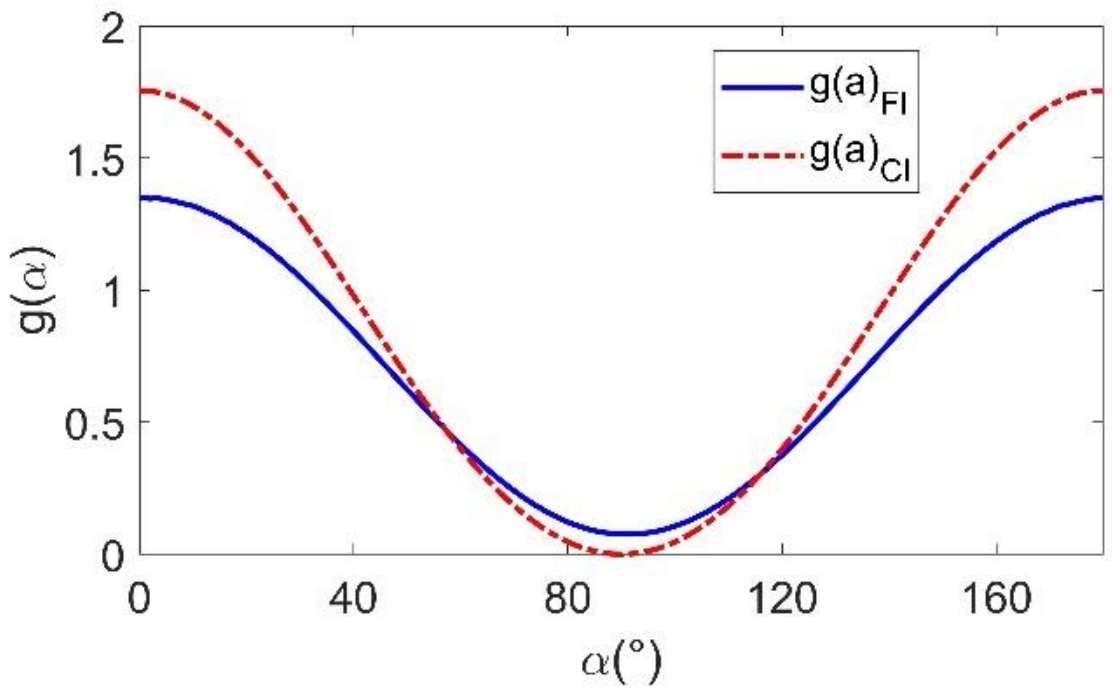

Figure 12. The angular distribution of atom pair F-I, shown by the blue curve, is measured experimentally. The retrieved angular distribution of C-I is the red dash curve. Angular distribution of C-I is the orientation distribution of $CF_3I$ molecules.

The angular distribution of C-I of $CF_3I$ molecules is required to calculate the diffraction patterns using the simple mode of dissociation. Here we extracted the angular distribution of F-I from the ΔMPDF in polar representation, which is used to retrieve the angular distribution of C-I. In our experiment, the $CF_3I$ molecules are impulsively aligned with the IR laser and subsequently are triggered by the UV laser. Part of the aligned $CF_3I$ molecules are excited by the UV laser to initiate dissociation. The 1/e rotational dephasing timescale of the parent $CF_3I$ molecule is 1.5 ps, which is evaluated in ref. [3] based on a free-jet expansion model [84]. For molecules with larger moments of inertia, the dephasing time is longer. Therefore, the orientation distribution of parent $CF_3I$ molecules, and product $CF_3$-I molecules with an extended C-I distance, are approximately fixed during the photodissociation over a temporal window of ~0.5 ps. In the experimental ΔMPDF, the F-I distance can be clearly isolated from other atom pairs. Now we can focus on the signal of F-I pair to extract its angular distribution using eqn (B9). The parent and product molecules are approximated to be not rotating over the measurement window, the change of the angular distribution of atom pair F-I is zero, formulated as $\Delta g_{FI}(\alpha) = 0$. At $t$ = 0.4 ps, the F-I distance is elongated, which is indicated by the positive signal between 3.5 to 4 in Figure 11(c), and is clearly isolated from the original distance $r_{\mathrm{FI}}$ =2.89 Å. Therefore, distribution of ΔMPDF at $r_{\mathrm{FI}}$ =2.89 Å is given by $-g_{FI}(\alpha)p(r - 2.89\text{ Å})$ . By integrating $-g_{FI}(\alpha)p(r - 2.89\text{ Å})$ along the $r$ axis around 2.89 Å, we can obtain the unnormalized $g_{FI}(\alpha)$. The normalized angular distribution of F-I is given by $g_{FI}(\alpha)/\int_0^{\pi} g_{FI}(\alpha)\sin\alpha\, d\alpha$ , and is shown as blue line in Figure 12. The C-I angular distribution $g_{CI}(\alpha)$, which is orientation distribution of $CF_3I$ molecules (shown as the red dash line in Figure 12), is retrieved using the method described in [51].

## APPENDIX G: $CF_3I$ A-BAND EXCITATION

The UV absorption of $CF_3I$ in the A-band consists of three repulsive states overlapped in energy denoted as $^3Q_1$, $^3Q_0$, and $^1Q_1$, involving promotion of a non-bonding (n) electron from the iodine atom valence shell to the σ* anti-bonding orbital of the C-I bond. The repulsive potential energy surfaces result in a prompt cleavage of the C-I bond [38-40, 42-45, 85] to produce excited and ground state iodine atoms. The $^3Q_0$ state excitation involves a transition dipole moment parallel to the C-I bond, and the corresponding photodissociation product is the spin-orbit excited state iodine $\mathrm{I}^*(^2\mathrm{P}_{1/2})$ [44, 86, 87]. The transition dipole moments for excitation of the $^3Q_1$ and $^1Q_1$ states are perpendicular to the C-I bond, producing ground state iodine $\mathrm{I}(^2\mathrm{P}_{3/2})$ [44, 86, 87]. Ground state $\mathrm{I}(^2\mathrm{P}_{3/2})$ can also be generated by the non-adiabatic coupling of the $^3Q_0$ and $^1Q_1$ states [43, 86]. The reported quantum yield of $\mathrm{I}^*(^2\mathrm{P}_{1/2})$ upon A-band excitation is ~0.9, which is almost one order of magnitude larger than the yield of $\mathrm{I}(^2\mathrm{P}_{3/2})$ [40, 65]. Therefore, the signal measured by GUED is mostly contributed from the dissociation channel that produces $\mathrm{I}^*(^2\mathrm{P}_{1/2})$. The energy to break CI bond of $CF_3I$ to produce $\mathrm{CF_3} + \mathrm{I}^*(^2\mathrm{P}_{1/2})$ is 2.5 eV, and to produce $\mathrm{CF_3} + \mathrm{I}(^2\mathrm{P}_{3/2})$ is 3.5 eV [45].

## APPENDIX H: DIFFRACTION PATTERN SIMULATION AND MAPPING DISTANCE TO TIME

We use a simple model to simulate the dissociation process (the model is described in the main text in section IV.C) to validate our interpretation of the $\Delta\mathrm{PDF}_{\parallel}$ signal that captures the dissociation dynamics. In order to simulate the MPDF, the orientation distribution of the ensemble of excited $CF_3I$ molecules $g_{CI}(\alpha)$ is needed. This was determined in APPENDIX F for the case of IR+UV excitation. For the UV-only case, the $g_{CI}(\alpha)$ is assumed to be $\frac{2}{\pi}\cos^2\alpha$ [3, 53]. We use the fast calculation algorithm described in [83] to calculate the diffraction pattern for a series of different values of ΔCI from 0 Å to 15 Å. The difference pattern is calculated as a function of the distance change, $\Delta I_M = I_M(\boldsymbol{s}, \Delta\mathrm{CI}) - I_M(\boldsymbol{s}, \Delta\mathrm{CI} = 0)$.

The limitations of the experimental data, specifically the missing data at low $s$ values and residual background can potentially introduce artifacts in the $\Delta\mathrm{PDF}_{\parallel}$. We use our simulations not only to compare with the experimental data but also to carefully check whether the data analysis introduces artifacts. Here we compare the $\Delta\mathrm{PDF}_{\parallel}$ of the simulated diffraction patterns in two cases. The first case is that we only apply the Gauss damping to simulated $\Delta\mathcal{M}(\boldsymbol{s})$ and subsequently calculate the $\Delta$MPDF and $\Delta\mathrm{PDF}_{\parallel}$. The second case is that we apply the same procedures used in the experimental data analysis, including the interpolation of missing data region, background correction and Gauss damping, to calculate the $\Delta$MPDF and $\Delta\mathrm{PDF}_{\parallel}$. The calculated $\Delta\mathrm{PDF}_{\parallel}$ for the first and second cases are shown in Figure 13 (a) and (b), respectively. The comparison shows that artifacts appear in the $\Delta\mathrm{PDF}_{\parallel}$ for $r \leq 1.50$ Å due to the background correction and the interpolation of missing data. However, the features of the $\Delta\mathrm{PDF}_{\parallel}$ that we use for our analysis are all at $r > 1.50$ Å, where the signal is not affected. Therefore, we expect the signal of C-I and F-I elongation to be retrieved accurately.

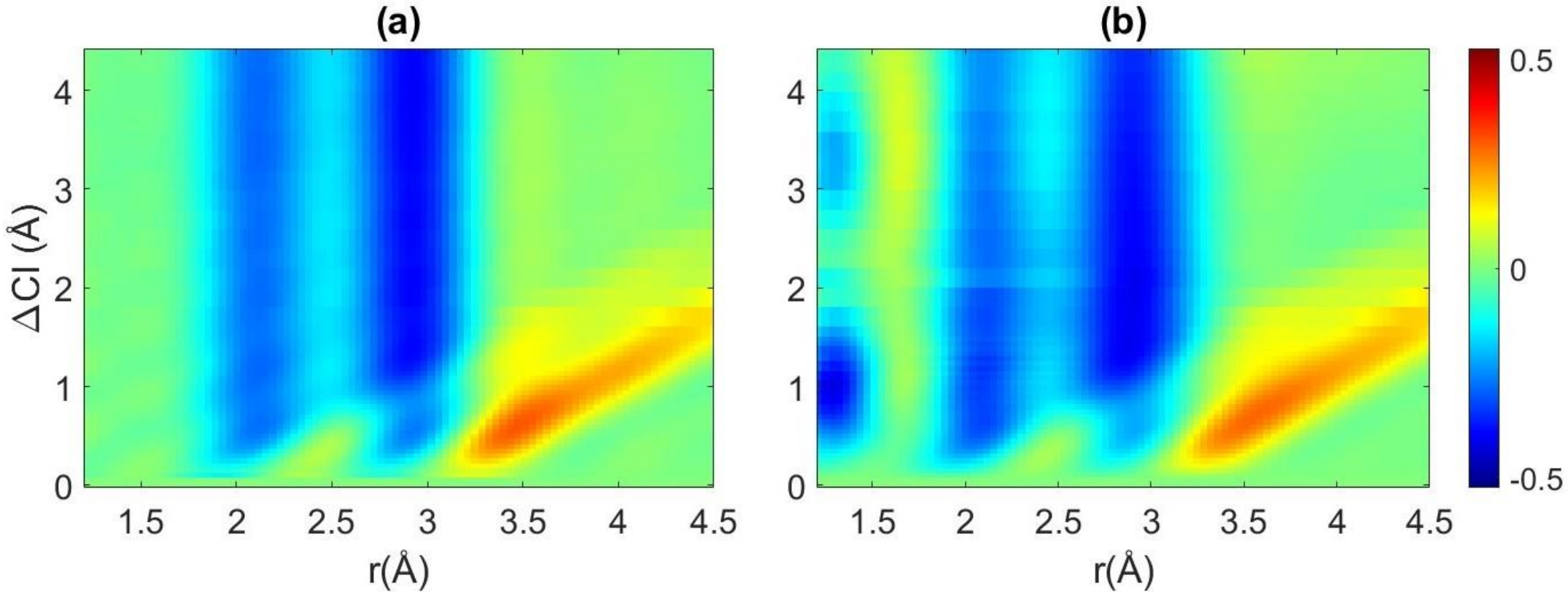

Figure 13. Simulated $\Delta\mathrm{PDF}_{\parallel}(\mathrm{r}, \Delta\mathrm{CI})$. (a) Calculated $\Delta\mathrm{PDF}_{\parallel}$ with only Gauss damping applied. (b) Calculated $\Delta\mathrm{PDF}_{\parallel}$ with the same treatment of the experimental data applied to the simulated $\Delta\mathcal{M}(\boldsymbol{s}, t)$.

The next step is to convert the $\Delta$CI to time for comparing with the experimental data. For this, we use the dissociation speed extracted from the experimental $\Delta\mathrm{PDF}_{\parallel}$ for the IR+UV case, and for the UV only excitation we use a value of the speed from the literature. The experimental $\Delta\mathrm{PDF}_{\parallel}(\mathrm{r}, \mathrm{t})$ for the IR+UV experiments is shown in Figure 6 (b) main text. The changing F-I distance, indicated by the positive tilted signal, can be clearly identified and used to extract the relative speed of the fragments. We first retrieve the peak position of the F-I signal at each time delay. We then use a linear fit to extract the speed of the moving wavepacket, which gives a value of 2.37± 0.23 Å/ps. The simulated $\Delta\mathrm{PDF}_{\parallel}$ based on the simple model is a function of $\Delta$CI, shown in Figure 14.

For IR+UV excitation, we converted the $\Delta$CI (vertical axis of Figure 14) to time (ps) by comparing the slope of the tilted F-I signal in the experimental $\Delta\mathrm{PDF}_{\parallel}(\mathrm{r}, \mathrm{t})$ to the counterpart in the simulated $\Delta\mathrm{PDF}_{\parallel}(r, \Delta\mathrm{CI})$ shown in Figure 14. The peak positions of F-I signal at all $\Delta$CI are obtained in unit of Å, which are fitted to a linear function to obtain the slope of the titled F-I signal. The fitted slope is 1.19± 0.04, and the red line in Figure 14 is the linear function with fitted parameters. Therefore, the ratio of converting $\Delta$CI (Å) to time (ps) is 2.82 ± 0.29 Å/ps, which is the rate of C-I elongation, called dissociation speed in the main text.

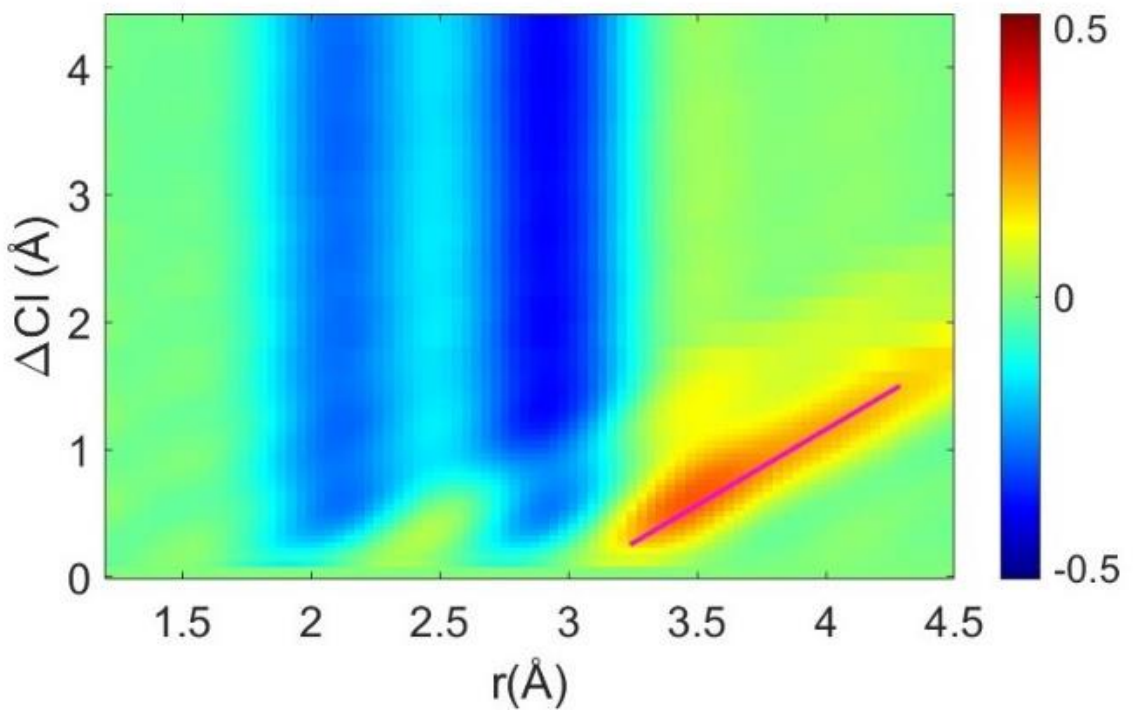

Figure 14. Simulated $\Delta\mathrm{PDF}_{\parallel}(r, \Delta\mathrm{CI})$ based on the simple model. The peak positions of F-I signal at all $\Delta$CI are obtained and fitted to a linear function. The fitted slope is 1.19± 0.04, and the red line is the linear function with fitted parameters.

After converting the simulated $\Delta\mathrm{PDF}_{\parallel}$ to time, it is convolved with a Gaussian function with a FWHM (150 fs) to match the IRF of the experiment. The calculated $\Delta\mathrm{PDF}_{\parallel}$ without convolution is shown in Figure 15 (a), and it is shown after the convolution in Figure 15 (b) and Figure 7(b) in the main text. The convolution affects mostly the data at early times and smears the positive peak from the dissociation wavepacket.

We also applied the same procedure to simulate $\Delta\text{PDF}_{\parallel}$ for the UV-only experiment. We used a dissociation speed of 22.51 Å/ps retrieved from the *ab-initio* calculation for one UV photon excitation of $CF_3I$ reported in [63]. The calculated $\Delta\text{PDF}_{\parallel}$ without and with convolution for UV excitation are shown in Figure 16 (a) and (b). Here, because of the higher speed of dissociation, the convolution completely blurs out the positive signal from the dissociating wavepacket.

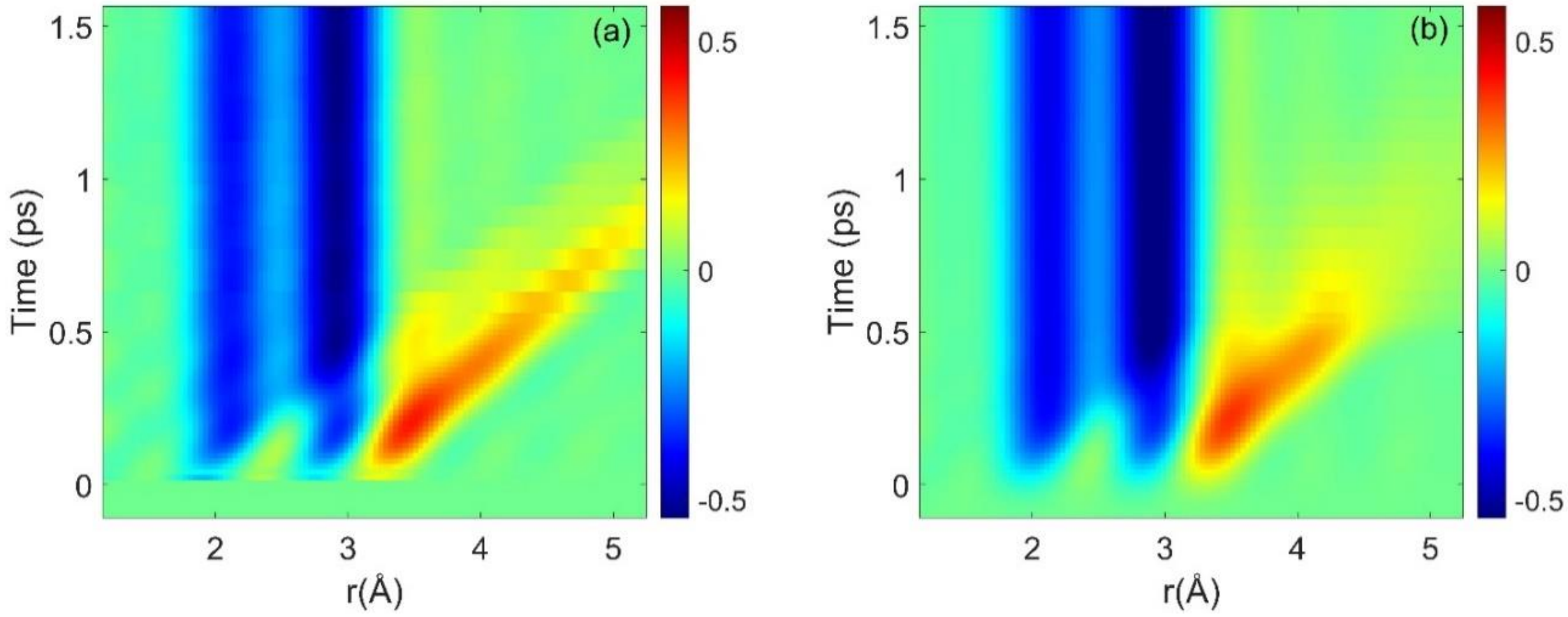

Figure 15. Simulated $\Delta\text{PDF}_{\parallel}(\text{r}, \text{t})$ for IR+UV excitation: (a) without convolution, and (b) convolving with a Gaussian function with a FWHM width of 150 fs that reflects the IRF.

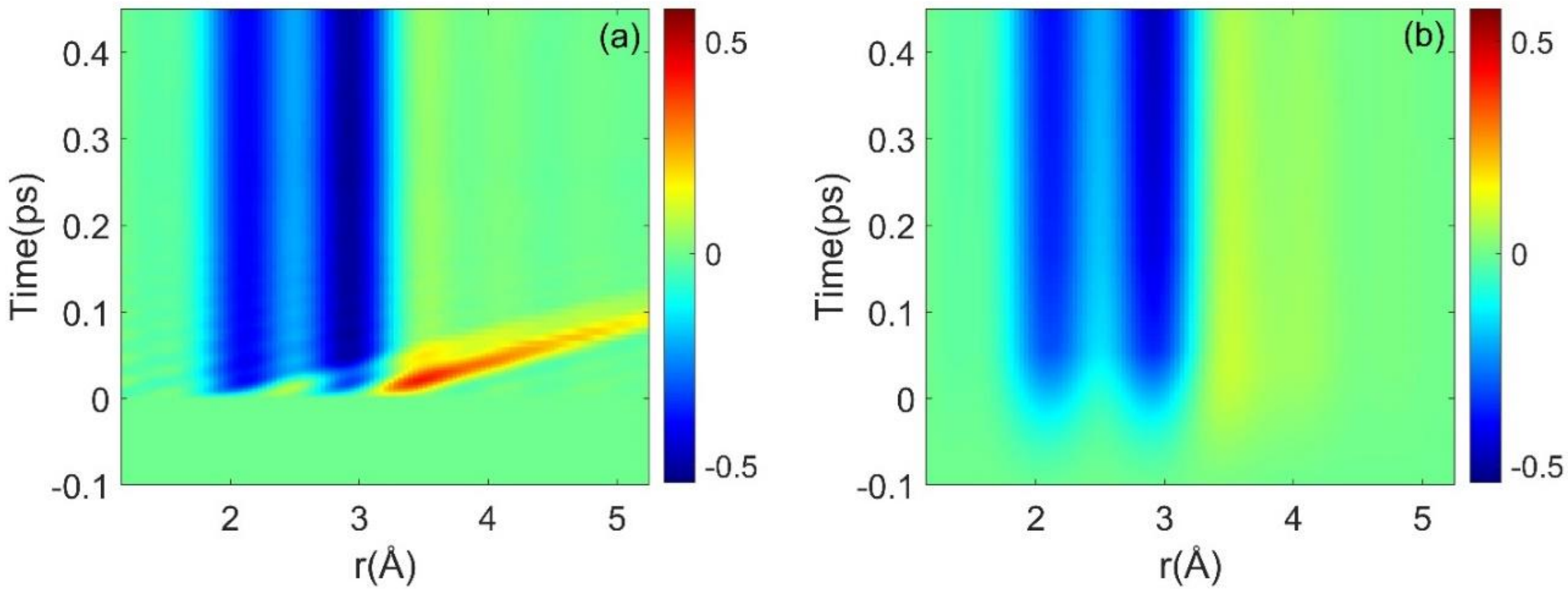

Figure 16. Simulated $\Delta\text{PDF}_{\parallel}(\text{r}, \text{t})$ for UV-only excitation: (a) without convolution, and (b) convolving with a Gaussian function with a FWHM width of 150 fs that reflects the IRF.

## APPENDIX I: COMBINATION OF FAST AND SLOW SIGNALS

The data for the IR+UV excitation shown in Figure 6(b) main text is best matched by a combination of fast and slow channels. This is quantified by calculating the ratio (absolute value) of the amplitude of the tilted F-I distribution and F-I depletion signal, and compare the ratios obtained from main text Figure 6(b) and Figure 7(b) that corresponds to signal of 100% slow reaction channel to estimate the relative yield of the two channels. For example, the ratio of the amplitude of the tilted F-I distribution to F-I depletion signal in Figure 6 (b) at $t$ = 0.3 ps is 0.32, whereas the ratio at the same time delay is 0.84 obtained from Figure 7(b). Therefore, the yield of the slow reaction channel is 0.37 by comparing the signals at $t$ = 0.3 ps. We compared the ratios obtained from Figure 6(b) and Figure 7(b) from 0.2 ps to 0.4 ps, and the relative yield of the slow reaction channel is estimated to be 0.40 ± 0.05. The relative yield of the slow reaction channel is estimated to be 0.4, and that of the fast reaction channel is 0.6. Therefore, the total signal that contains both the slow and fast reaction channels is given by $0.4 \times \Delta\text{PDF}_{\parallel}{}^{slow} + 0.6 \times \Delta\text{PDF}_{\parallel}{}^{fast}$ , where $\Delta PDF_{\parallel}{}^{fast}$ and $\Delta PDF_{\parallel}{}^{slow}$ are shown in Figure 7 (a) and (b) main text. The total signal, which is shown in Figure 17, matches the signal amplitude of experimental result in Figure 6(b) main text. However, since the simple model does not consider the vibrational motion of $CF_3$, the early dynamics in modeled total signal shown in Figure 17 do not resemble the counterpart in experimental result.

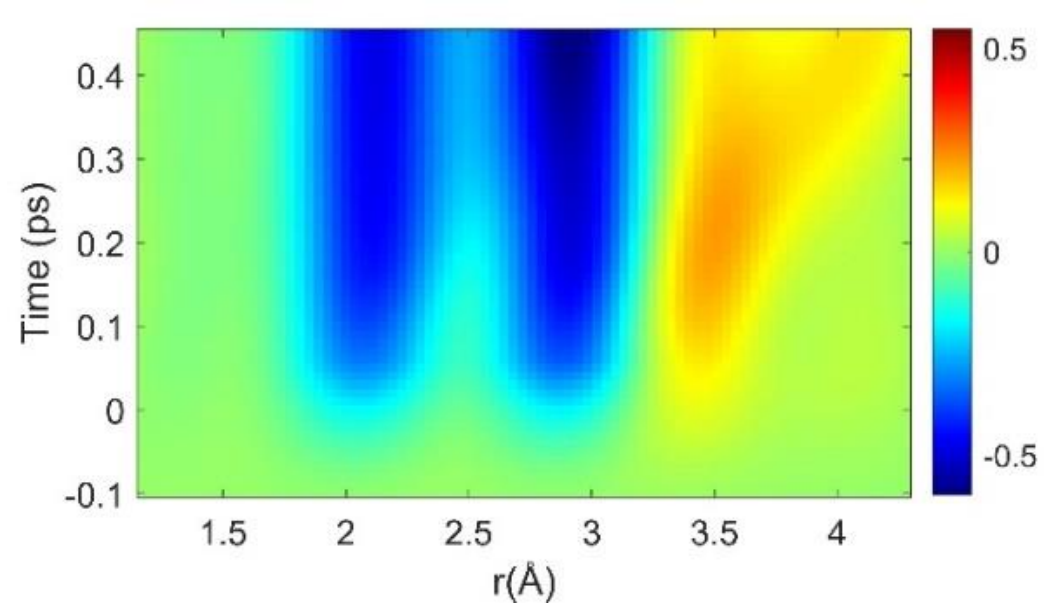

Figure 17. The total signal contains both the slow and fast reaction channels.

1. Hensley, C.J., J. Yang, and M. Centurion, *Imaging of isolated molecules with ultrafast electron pulses.* Physical review letters, 2012. **109**(13): p. 7035-7040.
2. Yang, J., et al., *Diffractive Imaging of Coherent Nuclear Motion in Isolated Molecules.* Phys Rev Lett, 2016. **117**(15): p. 153002.
3. Yang, J., et al., *Imaging CF3I conical intersection and photodissociation dynamics with ultrafast electron diffraction.* Science, 2018. **361**(6397): p. 64-67.
4. Liu, Y., et al., *Spectroscopic and Structural Probing of Excited-State Molecular Dynamics with Time-Resolved Photoelectron Spectroscopy and Ultrafast Electron Diffraction.* Physical Review X, 2020. **10**(2).
5. Champenois, E.G., et al., *Femtosecond Electronic and Hydrogen Structural Dynamics in Ammonia Imaged with Ultrafast Electron Diffraction.* Phys Rev Lett, 2023. **131**(14): p. 143001.
6. Centurion, M., T.J.A. Wolf, and J. Yang, *Ultrafast Imaging of Molecules with Electron Diffraction.* Annu Rev Phys Chem, 2022. **73**: p. 21-42.
7. Srinivasan, R., et al., *Dark structures in molecular radiationless transitions determined by ultrafast diffraction.* Science, 2005. **307**(5709): p. 558-63.
8. Ihee, H., et al., *Direct imaging of transient molecular structures with ultrafast diffraction.* Science, 2001. **291**(5503): p. 458-62.
9. Weathersby, S., et al., *Mega-electron-volt ultrafast electron diffraction at SLAC National Accelerator Laboratory.* Review of Scientific Instruments, 2015. **86**(7): p. 073702.
10. Shen, X., et al., *Femtosecond gas-phase mega-electron-volt ultrafast electron diffraction.* Struct Dyn, 2019. **6**(5): p. 054305.
11. Filippetto, D., et al., *Ultrafast electron diffraction: Visualizing dynamic states of matter.* Reviews of Modern Physics, 2022. **94**(4): p. 045004.
12. Ma, Z., et al., *Ultrafast isolated molecule imaging without crystallization.* Proc Natl Acad Sci U S A, 2022. **119**(15): p. e2122793119.
13. Jiang, H., et al., *Super-resolution femtosecond electron diffraction reveals electronic and nuclear dynamics at conical intersections.* Nat Commun, 2025. **16**(1): p. 6703.
14. van Oudheusden, T., et al., *Compression of subrelativistic space-charge-dominated electron bunches for single-shot femtosecond electron diffraction.* Phys Rev Lett, 2010. **105**(26): p. 264801.
15. Chatelain, R.P., et al., *Ultrafast electron diffraction with radio-frequency compressed electron pulses.* Applied Physics Letters, 2012. **101**(8).
16. Xiong, Y., K.J. Wilkin, and M. Centurion, *High-resolution movies of molecular rotational dynamics captured with ultrafast electron diffraction.* Physical Review Research, 2020. **2**(4).
17. Wilkin, K.J., et al., *Ultrafast electron diffraction from transiently aligned asymmetric top molecules: Rotational dynamics and structure retrieval.* Structural Dynamics, 2022. **9**(5): p. 054303.
18. Yang, J., et al., *Simultaneous observation of nuclear and electronic dynamics by ultrafast electron diffraction.* Science, 2020. **368**(6493): p. 885-889.
19. Yang, J., et al., *Diffractive imaging of a rotational wavepacket in nitrogen molecules with femtosecond megaelectronvolt electron pulses.* Nature communications, 2016. **7**.
20. Yang, J., et al., *Diffractive Imaging of Coherent Nuclear Motion in Isolated*

*Molecules.* Physical Review Letters, 2016. **117**(15): p. 153002.
21. Champenois, E.G., et al., *Conformer-specific photochemistry imaged in real space and time.* Science, 2021. **374**(6564): p. 178-182.
22. Liu, Y., et al., *Rehybridization dynamics into the pericyclic minimum of an electrocyclic reaction imaged in real-time.* Nature Communications, 2023. **14**(1): p. 2795.
23. Figueira Nunes, J.P., et al., *Monitoring the evolution of relative product populations at early times during a photochemical reaction.* Journal of the American Chemical Society, 2024. **146**(6): p. 4134-4143.
24. Muvva, S.B., et al., *Ultrafast structural dynamics of UV photoexcited cis, cis-1, 3-cyclooctadiene observed with time-resolved electron diffraction.* Physical Chemistry Chemical Physics, 2025. **27**(1): p. 471-480.
25. Nunes, J.P.F., et al., *Photo-induced structural dynamics of o-nitrophenol by ultrafast electron diffraction.* Physical Chemistry Chemical Physics, 2024. **26**(26): p. 17991-17998.
26. Hensley, C.J., J. Yang, and M. Centurion, *Imaging of isolated molecules with ultrafast electron pulses.* Physical review letters, 2012. **109**(13): p. 133202.
27. Xiong, Y., et al., *Isotope detection in molecules with ultrafast electron diffraction and rotational spectrometry.* Journal of Physics Communications, 2022.
28. Xiong, Y., et al., *Strong-field induced fragmentation and isomerization of toluene probed by ultrafast femtosecond electron diffraction and mass spectrometry.* Faraday Discuss, 2021.
29. Heo, J., et al., *Capturing the generation and structural transformations of molecular ions.* Nature, 2024. **625**(7996): p. 710-714.
30. Yang, J., et al., *Structure retrieval in liquid-phase electron scattering.* Physical Chemistry Chemical Physics, 2021. **23**(2): p. 1308-1316.
31. Nunes, J.P.F., et al., *Liquid-phase mega-electron-volt ultrafast electron diffraction.* Struct Dyn, 2020. **7**(2): p. 024301.
32. Yang, J., et al., *Direct observation of ultrafast hydrogen bond strengthening in liquid water.* Nature, 2021. **596**(7873): p. 531-535.
33. Townsend, D., B.J. Sussman, and A. Stolow, *A Stark Future for Quantum Control.* The Journal of Physical Chemistry A, 2011. **115**(4): p. 357-373.
34. Rosenwaks, S., *Vibrationally Mediated Photodissociation*. 2009: Royal Society of Chemistry.
35. Likar, M.D., et al., *Vibrationally mediated photodissociation.* Journal of the Chemical Society, Faraday Transactions 2: Molecular and Chemical Physics, 1988. **84**(9): p. 1483-1497.
36. Bisgaard, C.Z., et al., *Excited-state dynamics of isolated DNA bases: a case study of adenine.* Chemphyschem, 2009. **10**(1): p. 101-10.
37. Sussman, B.J., et al., *Dynamic stark control of photochemical processes.* Science, 2006. **314**(5797): p. 278-281.
38. Dzvonik, M., S. Yang, and R. Bersohn, *Photodissociation of molecular beams of aryl halides.* The Journal of Chemical Physics, 1974. **61**(11): p. 4408-4421.
39. Durie, R.A., T. Iredale, and J.M.S. Jarvie, *Absorption spectra of substances containing the carbon-iodine bond. Part I.* Journal of the Chemical Society (Resumed), 1950: p. 1181-1184.
40. Van Veen, G.N.A., et al., *Photofragmentation of CF3I in the A band.* Chemical Physics, 1985. **93**(2): p. 277-291.
41. Cheng, P.Y., D. Zhong, and A.H. Zewail, *Kinetic-energy, femtosecond resolved reaction dynamics. Modes of dissociation (in iodobenzene) from time-velocity correlations.* Chemical Physics Letters, 1995. **237**(5): p. 399-405.
42. Sage, A.G., et al., *nsigma* and pisigma* excited states in aryl halide photochemistry: a comprehensive study of the UV photodissociation dynamics of iodobenzene.* Phys Chem Chem Phys, 2011. **13**(18): p. 8075-93.
43. Zhang, X.P., W.B. Lee, and K.C. Lin, *Nonadiabatic transition in the A-band photodissociation of ethyl iodide from 294 to 308 nm by using velocity imaging detection.* J Phys Chem A, 2009. **113**(1): p. 35-9.
44. Mulliken, R.S., *Intensities in Molecular Electronic Spectra X. Calculations on Mixed-Halogen, Hydrogen Halide, Alkyl Halide, and Hydroxyl Spectra.* The Journal of Chemical Physics, 1940. **8**(5): p. 382-395.
45. Yin, S., et al., *Femtosecond pump–probe mass spectra on the dissociative photoionization of CF3I.* Chemical Physics Letters, 2003. **372**(5): p. 904-910.
46. Zandi, O., et al., *High current table-top setup for femtosecond gas electron diffraction.* Structural Dynamics, 2017. **4**(4): p. 044022.

47. Wang, Y., et al., *Ultrafast electron diffraction instrument for gas and condensed matter samples.* Rev Sci Instrum, 2023. **94**(5).
48. Charles Williamson, J. and A.H. Zewail, *Ultrafast electron diffraction. Velocity mismatch and temporal resolution in crossed-beam experiments.* Chemical Physics Letters, 1993. **209**(1): p. 10-16.
49. Baum, P. and A.H. Zewail, *Breaking resolution limits in ultrafast electron diffraction and microscopy.* Proceedings of the National Academy of Sciences, 2006. **103**(44): p. 16105-16110.
50. Zhang, P., J. Yang, and M. Centurion, *Tilted femtosecond pulses for velocity matching in gas-phase ultrafast electron diffraction.* New Journal of Physics, 2014. **16**(8): p. 083008.
51. Xiong, Y., et al., *Retrieval of the molecular orientation distribution from atom-pair angular distributions.* Physical Review A, 2022. **106**(3).
52. Ihee, H., et al., *Ultrafast Electron Diffraction and Structural Dynamics: Transient Intermediates in the Elimination Reaction of C2F4I2.* The Journal of Physical Chemistry A, 2002. **106**(16): p. 4087-4103.
53. Baskin, J.S. and A.H. Zewail, *Oriented ensembles in ultrafast electron diffraction.* Chemphyschem, 2006. **7**(7): p. 1562-74.
54. Glownia, J.M., et al., *Self-Referenced Coherent Diffraction X-Ray Movie of Angstrom- and Femtosecond-Scale Atomic Motion.* Phys Rev Lett, 2016. **117**(15): p. 153003.
55. Seideman, T., *Rotational excitation and molecular alignment in intense laser fields.* The Journal of Chemical Physics, 1995. **103**(18): p. 7887-7896.
56. Seideman, T. and E. Hamilton, *Nonadiabatic alignment by intense pulses. Concepts, theory, and directions.* ADVANCES IN ATOMIC, MOLECULAR AND OPTICAL PHYSICS, 2006.
57. Friedrich, B. and D. Herschbach, *Alignment and Trapping of Molecules in Intense Laser Fields.* Physical review letters, 1995. **74**(23): p. 4623-4626.
58. Larsen, J.J., et al., *Aligning molecules with intense nonresonant laser fields.* The Journal of Chemical Physics, 1999. **111**(17): p. 7774-7781.
59. Bar, I., et al., *Direct observation of preferential bond fission by excitation of a vibrational fundamental: Photodissociation of HOD (0,0,1).* The Journal of Chemical Physics, 1990. **93**(3): p. 2146-2148.
60. Bar, I., et al., *Mode-selective bond fission: Comparison between the photodissociation of HOD (0, 0, 1) and HOD (1, 0, 0).* The Journal of chemical physics, 1991. **95**(5): p. 3341-3346.
61. David, D., et al., *State-to-state photodissociation of the fundamental symmetric stretch vibration of water prepared by stimulated Raman excitation.* The Journal of Chemical Physics, 1993. **98**(1): p. 409-419.
62. Xiong, Y. and M. Centurion, *Fast calculation of diffraction patterns from an ensemble of aligned molecules.* Physical Review A, 2025. **112**(1).
63. Yang, J., et al., *Imaging CF3I conical intersection and photodissociation dynamics by ultrafast electron diffraction*. 2018, figshare.
64. Lin, D., et al., *Resolved (v1, v2 = 1) Combination Vibrational States of CF3 Fragments in the Photofragment Translational Spectra of CF3I.* The Journal of Physical Chemistry A, 2016. **120**(49): p. 9682-9689.
65. Suh, M., et al., *Energy Relaxation Dynamics of Photofragments Measured by Probe Beam Deflection Technique: Photodissociation of CF3I at 266 nm.* The Journal of Physical Chemistry A, 1999. **103**(42): p. 8365-8371.
66. Bar, I., et al., *Direct observation of preferential bond fission by excitation of a vibrational fundamental: Photodissociation of HOD (0, 0, 1).* The Journal of Chemical Physics, 1990. **93**(3): p. 2146-2148.
67. Crim, F.F., *Vibrationally Mediated Photodissociation: Exploring Excited-State Surfaces and Controlling Decomposition Pathways.* Annual Review of Physical Chemistry, 1993. **44**(1): p. 397-428.
68. Zittel, P. and D. Little, *Photodissociation of vibrationally excited hydrogen bromide.* The Journal of Chemical Physics, 1979. **71**(2): p. 713-722.
69. Zittel, P. and D. Little, *Photodissociation of vibrationally excited ozone.* The Journal of Chemical Physics, 1980. **72**(11): p. 5900-5905.
70. Zittel, P. and D. Little, *Photodissociation of vibrationally excited CH3Br.* Chemical Physics, 1981. **63**(1-2): p. 227-236.
71. Zittel, P. and L. Darnton, *Sulfur-33 isotope enrichment by isotopically selective, two-step, laser photodissociation of OCS.* The Journal of Chemical Physics, 1982. **77**(7): p. 3464-3470.

72. Zittel, P., L. Darnton, and D. Little, *Separation of O, C, and S isotopes by two-step, laser photodissociation of OCS.* The Journal of chemical physics, 1983. **79**(12): p. 5991-6005.
73. Zittel, P. and D. Masturzo, *Photodissociation of vibrationally excited OCS.* The Journal of chemical physics, 1986. **85**(8): p. 4362-4372.
74. Zittel, P.F. and V.I. Lang, *Isotopically selective photodissociation of gas phase OCS at low termperatures.* Journal of Photochemistry and Photobiology A: Chemistry, 1991. **56**(2): p. 149-158.
75. Centurion, M., et al., *Picosecond electron deflectometry of optical-field ionized plasmas.* Nature Photonics, 2008. **2**(5): p. 315-318.
76. Centurion, M., et al., *Picosecond imaging of low-density plasmas by electron deflectometry.* Optics Letters, 2009. **34**(4): p. 539-541.
77. Yang, J., et al., *Femtosecond gas phase electron diffraction with MeV electrons.* Faraday Discussions, 2016. **194**(0): p. 563-581.
78. Xiong, Y., et al., *Data for ultrafast electron diffractive imaging of the dissociation of pre-excited molecules*. 2025, figshare. https://doi.org/10.6084/m9.figshare.30114940
79. Kreier, D. and P. Baum, *Avoiding temporal distortions in tilted pulses.* Optics Letters, 2012. **37**(12): p. 2373-5.
80. Brockway, L.O., *Electron Diffraction by Gas Molecules.* Reviews of Modern Physics, 1936. **8**(3): p. 231-266.
81. Zare, R.N., *Angular Momentum Understanding spatial aspects in chemistry and physics.* 1988: p. page 77-81.
82. Herzberg, G., *Molecular spectra and molecular structure. vol 2: Infrared and Raman spectra of polyatomic molecules.* 1945: p. page 27-28.
83. Xiong, Y. and M. Centurion, *Fast calculation of diffraction patterns from an ensemble of aligned molecules.* Physical Review A, 2025.
84. Hagena, O.F., *Nucleation and growth of clusters in expanding nozzle flows.* Surface Science, 1981. **106**(1-3): p. 101-116.
85. Cheng, P.Y., D. Zhong, and A.H. Zewail, *Kinetic-energy, femtosecond resolved reaction dynamics. Modes of dissociation (in iodobenzene) from time-velocity correlations.* Chemical Physics Letters, 1995. **237**(5-6): p. 399-405.
86. Hancock, G., et al., *266 nm photolysis of CF3I and C2F5I studied by diode laser gain FM spectroscopy.* Phys Chem Chem Phys, 2007. **9**(18): p. 2234-9.
87. Gardiner, S.H., et al., *Dynamics of the A-band ultraviolet photodissociation of methyl iodide and ethyl iodide via velocity-map imaging with 'universal' detection.* Phys Chem Chem Phys, 2015. **17**(6): p. 4096-106.